\documentstyle[revtex,eqsecnum]{aps}
\begin{document}
\draft
\begin{title}
Phase Diagram for a Luttinger Liquid coupled to Phonons in one Dimension\\
\end{title}
\author{Thierry Martin $^{1*}$ and Daniel Loss $^{2}$}
\begin{instit}
$^{1}$ Theoretical Division, CNLS, Los Alamos National Laboratory,
Los Alamos, NM 87545

$^{2}$ Department of Physics, Simon Fraser University,
Burnaby, BC, Canada V5A1S6
\end{instit}
\begin{abstract}
We consider a one--dimensional system consisting of electrons with
short--ranged repulsive
interactions and coupled to  small--momentum transfer acoustic phonons.
The interacting electrons are bosonized and described in terms of a
Luttinger liquid
which allows to calculate exactly the one-- and two--electron Green
function.
For non--interacting electrons,
the coupling to phonons {\it alone} induces a singularity
at the Fermi surface which is analogous to that
encountered for electrons with an instantaneous
attractive interaction. The exponents which
determine the presence of singlet/triplet superconducting
pairing fluctuations, and spin/charge density wave fluctuations
are strongly affected by the presence of the Wentzel--Bardeen
singularity, resulting in the favoring of superconducting fluctuations.
For the Hubbard
model the equivalent of a phase diagram is established,
as a function of: the electron--phonon coupling, the
electron filling factor, and the on site repulsion
between electrons. The Wentzel--Bardeen singularity
can be reached for arbitrary values of the
electron--phonon coupling constant by varying
the filling factor. This provides an effective mechanism
to push the system from the antiferromagnetic  into the metallic
phase, and finally into the superconducting phase as
the electron filling factor is increased towards half filling.
\end{abstract}
\narrowtext
\section{Introduction}
\label{introduction}

In the light of recent technological advances in nanostructure
fabrication, there has been a renewed interest in the properties of
one--dimensional electron systems. A two--dimensional electron gas
created in a GaAS/AlGaAs heterostructure can be confined in its
lateral
dimension by means of metallic gates, yielding to a good
approximation
a one--dimensional wire. Up to recently, the theoretical study
of quasi--one--dimensional mesoscopic systems \cite{Imry}
has focused
mostly on non--interacting electrons.
In the present work, we will consider
the properties of a system of electrons in one dimension
interacting via a short ranged repulsive
potential, which in turn interacts with phonons
through a deformation potential coupling.
The coupling to the phonons
gives rise to a retarded {\it attractive}
electron--electron interaction which
opens the possibility for superconducting fluctuations.
On the other hand, the instantaneous repulsion between electrons
may be so strong that only spin density wave (SDW) and charge density
wave (CDW) fluctuations can survive.
In addition, the type of fluctuations which dominate will
depend crucially on the electron filling factor.
We will show that
the interplay between the phonon--mediated attractive interaction
and the instantaneous repulsive interaction can be
studied systematically in one dimension,
and thus obtaining the equivalent of a phase diagram for the fluctuations.

Aside from the experimental motivations mentioned above,
one--dimensional models can provide useful
answers: they are in general easier to handle than their
higher dimensional analogues, for which
a full treatment of the correlation effects
is often difficult. While the one--dimensionality strictly
rules out the possibility of long range
order, it is still legitimate to ask what
type of ordering fluctuations exist in such systems,
such as singlet and triplet pairing fluctuations, or spin density wave
and charge density wave  fluctuations. This question
was addressed for a purely instantaneous interaction between electrons
in the seventies
\cite{Luther,Emery,Solyom}, and later on for non--interacting
electrons coupled to phonons \cite{Hirsh}.
More recently, the attention has focused
on interacting electrons coupled to phonons
\cite{Voit0,Voit1,Zimanyi,Voit2}.

The interest in the latter systems is by no means new: static properties of
1D
interacting electron systems coupled to low energy acoustic phonons
were considered four decades ago by Wentzel \cite{Wentzel} and
Bardeen
\cite{Bardeen} as a possible candidate for the theory of
superconductivity. The electrons were described in terms of
Tomonaga
bosons \cite{Tomonaga}, and it was shown that for a critical value of
the electron--phonon coupling constant, the system becomes {\it
unstable}
and aquires a negative compressibility. The thermodynamic
quantities for the electron-phonon system (within the Einstein
model)
were studied near this singular point \cite{Engelsberg}, indicating
the
presence of a phase transition: the divergence of the specific heat is
accompanied by a collapse of the system induced by the strong
electron--phonon interaction.  We shall
refer to this point as the Wentzel--Bardeen singularity
in the course of this work. Nevertheless, the Wentzel--Bardeen
singularity was not taken too seriously in the context of
superconductivity, as the (strong) coupling needed to trigger
this dramatic behavior was judged to have unphysical values.

More recently, a system similar to that of ref. \cite{Engelsberg}
was examined, with the conclusion that the coupling to low
energy acoustic phonons can be neglected in most systems because
the effects associated with the phonons contain the small
parameter $c/v_F$, where $c$ is the speed of sound and
$v_F$ is the Fermi velocity. As a result, this type of coupling
ceased to receive much attention, and the interest shifted towards
higher momentum processes where  phonons transfer (or backscatter) an
electron
from one side of the Fermi surface to the other side
\cite{Voit1,Zimanyi,Voit2}:
$2k_F$ phonon processes. These processes are understood to
be at the origin of the Peierls instability.
Nevertheless, in the present work, we shall
come back to the case of low energy acoustic phonons, and will
point out that for strongly correlated electron systems,
this type of coupling  can have spectacular
effects as one approaches the Wenztel--Bardeen singularity.
For strongly correlated systems, the effects associated with the
phonons
do {\it not} depend on the small parameter $c/v_F$: rather,
the Fermi velocity $v_F$ is replaced by the charge velocity $u_\rho$ associated
with low energy particle--hole excitations close to the Fermi surface.
In contrast to $v_F$, $u_\rho$ can depend strongly on the electron
filling factor: for the Hubbard model, $u_\rho$ vanishes at half
filling whereas $v_F$ does not. Near half filling, the decay of $u_\rho$
is most dramatic for low values of the on site repulsion
parameter $U$, i.e. the system is extremely sensitive to the
filling factor.
As a consequence, the effects associated
with these low energy phonons can no
longer be dismissed as a small correction. Below, we will argue that:
1) the Wentzel--Bardeen singularity can be reached for arbitrary
non-zero value of the
electron--phonon coupling constant as one approaches half
filling in the
Hubbard model. 2) Near the Wentzel--Bardeen  singularity, CDW and SDW
fluctuations
are strongly suppressed, and the system is first pushed towards an
intermediate (metallic) phase, and eventually into the superconducting phase.

The low energy properties
of a one--dimensional electron system
can be treated rigorously using the so--called bosonization technique.
This approach has been applied recently to a variety of problems
in mesoscopic physics: the transport properties of quasi--one--
dimensional
interacting electrons systems \cite{Kane,Matveev}, the persistent
current
of a one--dimensional ring of interacting fermions \cite{Loss,Martin}
among others.
In the Luttinger liquid picture, the field operator
describing the electrons are characterized by
a charge field and a spin field, which refer to the two different
types of elementary excitations which are present in the system at low
temperatures.
In the Luttinger liquid approximation, these excitations are totally
decoupled, and each is characterized
by a separate velocity. Because of this, charge and spin
excitations can be localized at different points in space.
For the particular type of interaction considered here,
the coupling to the phonons occurs only with
the charge degrees of freedom. The collective excitations
describing the  charge--phonon dynamics are then characterized
by two distinct velocities, which depend on the
charge velocity, the sound velocity, and the electron--phonon coupling.
This property is reminiscent of the phenomenon of hybridization.
As a consequence, the electronic system coupled to phonons
is characterized by three distinct velocities.

The first part of this paper will be concerned with the calculation
of the single--electron Green function where the phonon part is
integrated out. The corresponding Green function for
interacting spinless fermions coupled to phonons has been
calculated in the early eighties within perturbation theory,
assuming a linear
spectrum near the Fermi surface \cite{Apostol}, and the results
were expanded in the ratio $c/v_F$.
Here, we shall use a functional integral
approach, where the phonons can
be
explicitly traced out
of the problem.  This is the equivalent of summing up
all phonon diagrams of the perturbation series \cite{Apostol}.
The results for the Green function can in turn be used to
determine the
the momentum distribution function of the electrons,
which is of the Fermi--Dirac
type for a free electron gas.
However, both the instantaneous interaction and the phonon--mediated
retarded interaction
contribute to drive the system away from the free electron
behavior. This is in analogy with
the early work on bosonization
which showed that for one--dimensional electron systems,
the momentum distribution function departs from the
Fermi-Dirac distribution for arbitrarily small coupling
\cite{Mattis}.

In a second part, we will focus on the
many body properties of the system.
Two--particle correlation functions
in Matsubara representation
can be computed with the same method,
yielding the long range and long
(imaginary) time behavior of the singlet and
triplet pairing correlation function,
as well as of the the SDW and CDW
correlation functions. For an uncorrelated fermion
system in the limit of vanishing temperature, arbitrarily small
phonon coupling leads to a divergence of the Fourier transform
of the pairing correlation
function at low momentum and low frequency,
which is precisely the signature for superconducting
fluctuations\cite{Luther}. This picture will be modified
when we include
repulsive interactions.
For an interacting electron system with a short range repulsive
potential
between electrons and with a filling factor between
zero and half filling, but not close to these values,
the onset of superconducting fluctuations correspond
in general to a finite value
of the
electron--phonon coupling constant. This threshold value of the
phonon
coupling is directly related to the parameters of the
electronic model which is considered. As an
example, we study the
Hubbard model, where the Luttinger liquid
parameters can be computed exactly. This
will allow us to study the instabilities of
the system for filling factors close to half filling, and
therefore to study the sensibility to the Wentzel--Bardeen
singularity.

Finally, we can apply our results to study the behavior of
two coupled electron system or alternatively a electron system
coupled to a hole system, with a local interaction.
The density--density coupling has then precisely the same
form as for electrons coupled to low energy
phonons, and consequently
there is the possibility that one system
may induce superconducting fluctuations in the other.
While it is possible that the inclusion of $2k_F$ phonons
could modify the behavior of the electron--phonon
system, for two coupled electron systems,
there is no such possibility: the
Fermi velocities of each mode are fixed
and distinct from each other, and there is no cancellation
of the fast varying components of the density operator.
Only the slow varying component of these operators
has to be taken into account,
and our model is ``exact'' in this sense.

The paper is organized as follows: in Sec. II,
the Hamiltonian of the coupled
electron--phonon system is described, and a brief review of
the bosonization
technique is presented. In Sec. III, we introduce the
functional integral formalism to calculate the thermodynamic
quantities.
Sec. IV is devoted to the calculation the single particle Green
function.
The ordering correlation functions are considered in Sec. V.
We illustrate our results with the Hubbard model in Sec. VI.
The relevance of our results for coupled  electron chains
are discussed in Sec. VII.
We conclude in Sec.
VIII.

\section{Description of the model}
\label{model section}

The starting point of our approach is a
system of interacting electrons
on a lattice coupled to acoustic phonons in one dimension, with
periodic boundary conditions.
The Hamiltonian  describing the acoustic phonons is given by:
\begin{equation}
H_{p}={1\over 2}\int d x~[\zeta^{-1}\Pi_d^2+ \zeta c^2 (\partial_x
d)^2]~,
\label{phonon}\end{equation}
where $d$ is the phonon field operator, $ \Pi_d$ is its canonical
conjugate, $\zeta$ is the mass density of the lattice and $c$ is the
speed of sound.

Let $\psi_s$ be the field operator for fermions with spin $s$. In the
deformation potential approximation, the electron--phonon coupling is
given by \cite{Fetter}:
\begin{equation}
H_{el-p}=g\sqrt{\pi\over 2}\sum_s\int d
x~\psi_s^{\dagger}\psi_s \partial_x d~,
\label{coupling}\end{equation}
where $g$ is the electron--phonon coupling constant (the numerical
factor
$\sqrt{\pi/2}$ is introduced to simplify later algebra).
As mentioned in the introduction, alternative
choices
for the phonon interaction are possible,
as is the case when one studies
conducting polymers \cite{T L M,S S H} or molecular crystals
\cite{Holstein}. However, for mesoscopic systems such as GaAs
heterostructures
in the metallic regime and
at low temperatures  ($T<1 K$), the acoustic phonons which are
coupled by the deformation potential to the electronic degrees of
freedom constitute
a realistic model of the lowest energy modes
of the system.

We now turn to the interacting electron system. It has been
shown \cite{Haldane} that the low energy and long wave length behavior
of a system
of spinless
fermions with short range interaction can be described by a
continuum Hamiltonian expressed in terms of
bosonic fields. A similar treatment exists for fermions {\it with} spin
\cite{Emery,Solyom}.
We restrict ourselves to correlated electron models
with nearest neighbor hopping and short--ranged
repulsion/attraction. The field operator describing
electrons with spin $s=\pm 1/2$ ($=\uparrow ,\downarrow$)
is decomposed into a right and left moving part:
\begin{equation}
\psi_s (x)=e^{i k_F x}\psi_{s+} (x)+e^{-i k_F x}\psi_{s-} (x)~,
\label{right left}\end{equation}
The right  (left) moving field operator $\psi_{s+}$  ($\psi_{s-}$)
is then expressed as an exponential of bosonic fields:
\begin{equation}
\psi_{s\pm} (x)={1\over \epsilon L}\exp (i\sqrt{\pi} (\pm\varphi_s
(x)-\theta_s(x)))
{}~,
\label{decomposition}\end{equation}
where $\theta_s(x)=\int^x dx^{\prime}\Pi_s (x^{\prime}))$,
and the bosonic field $\Pi_s$ is the canonical
conjugate of $\varphi_s$.
We then introduce the bosonic fields describing the charge and spin
fields:
\begin{mathletters}
\begin{eqnarray}
\varphi_\rho&=&{1\over
\sqrt{2}}(\varphi_\uparrow+\varphi_\downarrow)~,
\label{charge}\\
\varphi_\sigma&=&{1\over \sqrt{2}}(\varphi_\uparrow
-\varphi_\downarrow)~,
\label{spin}\end{eqnarray}\end{mathletters}
and their respective canonical conjugates $\Pi_\rho$ and
$\Pi_\sigma$.
With these definitions the electron density operator becomes
\cite{Shankar2}:
\begin{equation}
\sum_s\psi_s^\dagger\psi_s=\sqrt{2\over\pi}\partial_x\varphi_\rho
{}~,
\label{density}\end{equation}
where we have kept only the slow spatial variations of the density
operator.
Expressed in terms of the variables in Eqs. (\ref{charge}) and
(\ref{spin}),
and their canonical conjugate,
the Hamiltonian describing the interacting electron separates
into a charge and spin contribution.
Each contribution is of
the sine--Gordon type. Schulz \cite{Schulz,Schulz2} has argued that
these Hamiltonians can be replaced by ``free'', quadratic
Hamiltonians,
provided that the parameters of each contributions
are properly renormalized by the interactions.
The quadratic Hamiltonian of Ref. \cite{Schulz} is our starting point:
\begin{equation}
H_{el}=H_\rho+H_\sigma ~,
\label{charge and spin}
\end{equation}
with
\begin{equation}
H_\nu={1\over 2}\int d x~\Biggl[u_\nu K_\nu\Pi_\nu^2
+{u_\nu \over K_\nu}(\partial_x\varphi_\nu)^2\Biggr]~,
\label{nu Hamiltonian}\end{equation}
with $\nu=\rho, \sigma$.The free electron gas limit corresponds to
$K_\rho=K_\sigma=1$ and $u_\rho=u_\sigma =v_F$, the Fermi
velocity of non--interacting
electrons. Note that we have omitted an additional cosine term
\cite{Schulz}
in the spin channel: this marginally
irrelevant contribution will be discussed briefly in Sec. \ref{divergence
subsection}.

For the Hubbard model with on--site repulsion,
the justification for this ``free'' Hamiltonian \cite{Schulz}
is based on the fact that the large $U$ and the small $U$ limit
represent the same phase of the model, and there are
no singular points between these two limits.
Consequently the interaction renormalize  to the  ``fixed point''
$K_\sigma =1$.
The remaining parameters $K_\rho$, $u_\rho$ and
$u_\sigma$
of the Hamiltonian in Eq. (\ref{charge and spin}) can be obtained
by solving numerically the equations describing the ground
\cite{Lieb}
and excited states \cite{Coll}, as will be shown later in Sec.
\ref{Hubbard section}.
For other, more general models where an exact solution is not
available,
the parameters
$u_\rho$, $u_\sigma$, $K_\rho$, $K_\sigma$ can be obtained, for
example,
by finite size diagonalization for small systems.

The electron--phonon
coupling originates from the electrostatic interaction between the
electrons and the atoms of the solid. Consequently,
the phonons couple only to $H_\rho$, since it is the total charge
density
from up and down spins that matters.


\section{Lagrangian formulation}
\label{Lagrangian section}

To calculate thermodynamic quantities, we use a Lagrangian
formulation
\cite{Fradkin}
where the partition function is represented as a functional integral
over the phonon  (ph) displacement field and the bosonic fields
associated
with the charge and spin degrees of freedom.
For example, if we wish to calculate the thermodynamic average of
a functional
$f[{\hat \varphi}_\nu,{\hat \Pi}_\nu]$ which depends
on the conjugate operators ${\hat \varphi}_\nu$ and
${\hat \Pi}_\nu$, with $\nu=\rho,\sigma$,
we can express this average in terms of a functional integral as
follows
\begin{eqnarray}
<:f[{\hat \varphi}_\nu,{\hat \Pi_\nu]}:>=
Z^{-1}\int D\varphi_\nu D\Pi_\nu Dd D\Pi_d&&~\exp\{-
S_E[\varphi_\nu,\Pi_\nu ,d,\Pi_d]\}\nonumber\\
&\times&f[\varphi_\nu,\Pi_\nu]~,
\label{average}\end{eqnarray}
where
$Z$ is the partition function. The symbol $:f:$ denotes the normal
ordered form of f with the momentum operators brought to the
left of their conjugate field operators. We shall see below that
this type of normal ordering only leads to oscillatory phase
factors and does not
affect the asymptotic behavior (in the thermodynamic limit)
which is our main interest. These oscillatory terms, however, are
relevant
for finite size effects\cite{Loss}.
The Euclidean action $S_E$ can be decomposed as follows:
\begin{equation}
S_E[\varphi_\nu,\Pi_\nu,d,\Pi_d]=\int_0^{\beta}d\tau~\int_0^L d x~
[L_{ph}+L_{\rho}+L_{\sigma}+L_{\rho-ph}]~.
\end{equation}
Here, $\tau$ is the imaginary time coordinate ($\beta^{-1}=k_BT$).
The corresponding Lagrangian densities for the phonons, the charge
and spin degrees of freedom, and the coupling term are defined by:
\begin{mathletters}
\begin{eqnarray}
L_{ph}&=&{1\over 2 \zeta}\Pi_d^2
+{\zeta c^2\over 2}  (\partial_x d)^2-i\Pi_d\partial_\tau d
\label{Lagrangian p}\\
L_{\nu}&=&{u_\nu K_\nu\over 2}\Pi_\nu^2
+{u_\nu \over 2K_\nu}(\partial_x\varphi_\nu)^2
-i\Pi_\nu\partial_\tau\varphi_\nu
\label{Lagrangian nu}\\
L_{\rho-ph}&=&g  (\partial_x d) (\partial_x\varphi_\rho)~,
\label{Lagrangian coupling}
\end{eqnarray}
\end{mathletters}
with $\nu=\rho,\sigma$.
Note that the Lagrangian density describing the system is quadratic
in the
fields, and the functional integral of Eq.
 (\ref{average}) will be performed exactly. This is a direct
consequence
of the bosonization method, as the fermion density is linear
in $\partial_x\varphi_\rho$  (see Eq.  (\ref{density})).

In principle, the functional integrals of Eq.  (\ref{average})
implicitly contains summations over
the winding numbers associated with the zero-modes of the
fields\cite{Loss}.
These zero-modes, however,
shall be ignored here as we are calculating only bulk quantities, but
they
should be included in the calculation of any finite size effects.
Consequently, we shall omit in the Fourier sums below the $
k=0$
contribution (this constraint becomes irrelevant in the
thermodynamic limit).

The cross terms $-i\Pi_d\partial_\tau d$ and
$-i\Pi_\nu\partial_\tau\varphi_\nu$ in Eqs. (\ref{Lagrangian p})
and
(\ref{Lagrangian nu}) can be eliminated if we redefine
the canonical conjugates as follows:
\begin{eqnarray}
\Pi_d&\rightarrow &\Pi_d-i\zeta\partial_{\tau} d\nonumber\\
\Pi_\nu&\rightarrow&\Pi_\nu-i\partial_{\tau}\varphi_\nu/u_\nu
K_\nu
\nonumber\end{eqnarray}
The Lagrangian densities for the phonon,
charge, and spin degrees of freedom
then become:
\begin{eqnarray}
L_{ph}&=&{1\over 2\zeta}\Pi_d^2
+{\zeta c^2\over 2} (\partial_x d)^2+{\zeta\over 2}(\partial_\tau
d)^2\label{Lagrangian p 2}\\
L_{\nu}&=&{u_\nu K_\nu\over 2}\Pi_\nu^2+{u_\nu\over 2K_\nu}
(\partial_x\varphi_\nu)^2+{1\over 2u_\nu
K_\nu}(\partial_\tau\varphi_\nu)^2
{}~.\nonumber\\
\label{Lagrangian nu 2}\end{eqnarray}
Later on, in the calculation of the Green function and the correlation
functions, we will have to keep in mind this shift in the
canonical conjugate $\Pi_\nu$.

As we shall later be concerned with the calculation of one and two
particle
Green function for {\it electrons} only, we can integrate out the
phonon
degrees of freedom at this point.
We must therefore perform the following Gaussian
integral:
\begin{eqnarray}
I_1[\varphi_\rho]=\int Dd~\exp\biggl[-\int d{\bf x}~(&{\zeta\over 2}&
 d[- \partial_{\tau}^2- c^2\partial_x^2]d
\nonumber\\
&-&g d\partial_x^2\varphi_\rho )\biggr]~,
\label{i 1}\end{eqnarray}
where $d{\bf x}=d xd\tau$.
Denoting by $[-\partial_{\tau}^2-c^2\partial_x^2]^{-1}$
the inverse of the operator acting on the phonon field on the right
hand side
of Eq.  (\ref{i 1}), the functional integration over the phonon field
gives:
\begin{equation}
I_1[\varphi_\rho]=c_1\exp[{g^2\over {2\zeta}}\int d{\bf x} d{\bf
x}^{\prime}
 (\partial_x^2\varphi_\rho) ({\bf x})
[-\partial_{\tau}^2-c^2\partial_x^2]^{-1} ({\bf x},{\bf
x}^{\prime})]
 (\partial_{x^{\prime}}^2\varphi_\rho) ({\bf
x}^{\prime})~.\label{result i 1}
\end{equation}
where $c_1=[\det (-\partial_{\tau}^2-c^2\partial_x^2)]^{-
1/2}$.
We can calculate the propagator in Eq.  (\ref{result i 1})
by going to a Fourier representation for the charge field thereby
assuming periodic boundary conditions in space and imaginary time:
\begin{equation}
\varphi_\rho ({\bf x})={1\over {\beta L}}
\sum_{\bf k} e^{i{\bf k}\cdot{\bf x}}\varphi_\rho({\bf k})~,
\end{equation}
with ${\bf k}= (k,\omega)$, ${\bf x}= (x,\tau)$ and
${\bf k}\cdot{\bf x}=kx+\omega t$, and where the $k=0$
term corresponding to the zero-mode is to be omitted henceforth.
We then obtain the contribution for the effective action
for the charge degrees of freedom:
\begin{equation}
{\tilde S}[\varphi_\rho]={1\over 2\beta L}
\sum_{\bf k}D_\rho({\bf k})|\varphi_\rho ({\bf k})|^2~,
\label{Fourier i 1}
\end{equation}
where we have introduced the inverse propagator for the charge
field
$\varphi_\rho$:
\begin{equation}
D_\rho({\bf k})={1\over K_\rho u_\rho}\biggl(\omega^2+u_\rho^2
k^2
-{b^2 k^4\over \omega^2+c^2k^2}\biggr)~,
\label{propagator}\end{equation}
where $b=g\sqrt{K_\rho u_\rho /\hbar\zeta}$ (with $\hbar$'s
restored)
is a constant which has the
dimension of a velocity squared. The first two terms
of Eq. (\ref{propagator})
represent the contribution of the  uncoupled charge field. The
coupling
to phonons therefore creates an additional {\it quadratic}
contribution to the action, which contains all retardation effects:
a collective excitation of the interacting electron system
may create a phonon at time $t$
and reabsorb it at a later time $t^{\prime}$.
We note that $k^2/D_{\rho}(\bf k)$ is the effective charge density
propagator
in Fourier space.
This retarded interaction is always attractive, and will
be responsible for the formation of Cooper pairs.
We also note that this retarded interaction term enters with a {\it
minus}
sign irrespective of the sign of the coupling constant $g$.
As a result, an instability arises at zero-frequency and when
$u_\rho$
approaches $b/c$ (from above) because $D_{\rho}$ then approaches zero,
signaling an instability in the charge density propagator,
$k^2/D_\rho({\bf k})$,
towards long wave--length density fluctuations.
This instability has been known
for a long time and has first been recognized by
Wentzel\cite{Wentzel}
and later by Bardeen\cite{Bardeen}, its consequences for the
thermodynamic (static) properties of a non-interacting
system have been discussed in Ref.~\cite{Engelsberg}.
This instability will be of great importance in our following
discussion
of the {\it dynamic} properties  of the Luttinger liquid, in particular
with respect to its various possible phases characterized by
the retarded response functions.

Had we treated the electrons with conventional fermion operators
the integration of the phonon variables would have resulted in an
effective interaction which is {\it quartic} in the fermion operators.
Moreover, in problems where a single particle is coupled to a phonon
bath
\cite{Feynman}, the ``influence functional'' which results
from the functional integration of the bath variables is in general
quite
complicated. The particle is then coupled to an effective potential
which contains ``memory effects'' \cite{Caldeira,Thierry}, and one has
to take recourse to approximation methods to
understand how the coupling to the phonon bath affects the single
particle motion. This is not the case here: a
one--dimensional electron system (with or without interactions)
coupled to phonons can be treated exactly if
one uses the bosonization description for the charge and spin fields.


\section{Single particle properties}
\label{single section}

In this section we discuss the effects of the
electron--phonon interaction on the single-particle properties.
In particular, we show that an arbitrarily small coupling to phonons
immediately drives the system into a non-Fermi liquid state, even
in the absence of electron interactions.

\subsection{Green function}
\label{Greens subsection}

We proceed to calculate the single--particle Green function:
\begin{equation}
G_s(x,\tau)=-<T\psi_s(x,\tau)\psi_s^\dagger(0,0)>~,
\label{Greens 0}\end{equation}
where $T$ is the imaginary time ordering operator.
Using the decomposition of the fermion operators into right and
left moving components (Eq. (\ref{decomposition}),
the Green function becomes:
\begin{equation}
G_s({x,\tau})=
-\sum_{\alpha,\alpha^\prime=\pm}
e^{i\alpha k_Fx}
<T\psi_{s\alpha}(x,\tau)\psi_{s\alpha^\prime}^\dagger(0,0)>~,
\label{Greens}\end{equation}
where  $\psi_{s\alpha} (x,\tau)$ is expressed in terms of bosonic
variables
as in Eq.  (\ref{decomposition}). The diagonal contribution
$\alpha=\alpha^\prime$ turns out to be the dominant one.
We thus need to calculate the expression
\begin{eqnarray}
<T\psi_{s\alpha}(x,\tau)\psi_{s\alpha}^\dagger(0,0)>
={1\over \epsilon L}
<&e&^{-i\sqrt{\pi\over 2}
(\theta_\rho(x,\tau)-\theta_\rho(0,0))}e^{\alpha i\sqrt{\pi\over
2}(\varphi_\rho(x,\tau)-
\varphi_\rho(0,0))}
\nonumber\\
&e&^{-is_\sigma\sqrt{\pi\over 2}(\theta_\sigma(x,\tau)-
\theta_\sigma(0,0))}
e^{\alpha s_\sigma i\sqrt{\pi\over 2}(\varphi_\sigma(x,\tau)-
\varphi_\sigma(0,0))}>\nonumber\\
&\times&~g_n(x,\tau)~,
\label{Greens Boson}
\end{eqnarray}
with the charge and spin fields introduced in Eqs. (\ref{charge})
and (\ref{spin}), and with $s_\rho=1$, and $s_\sigma=s$.
The last line on the r.h.s. of Eq. (\ref{Greens Boson})
is a c--number which
arises from the normal ordering procedure, and, moreover,
contains the information
about the time ordering of the Fermi operators:
\begin{eqnarray}
g_n(x,\tau)&=&+\prod_{\nu=\rho,\sigma}
e^{\alpha \pi \left[\theta_{\nu}(0,0),\varphi_{\nu}(x,\tau)\right]/2},
{}~~~{\rm for}~~~\tau>0
\nonumber\\
g_n(x,\tau)&=&-\prod_{\nu=\rho,\sigma}
e^{\alpha  \pi
\left[\theta_\nu(x,\tau),\varphi_\nu(0,0)\right]/2},~~~{\rm
for}~~~\tau<0~.\nonumber\\
\label{normal order Greens}\end{eqnarray}
The normal ordering contribution, Eq. (\ref{Greens Boson}),
involves the  commutators between $\theta_\nu$ and $\varphi_\nu$
evaluated at
unequal times. The calculation of these commutators is given in
Appendix A, where also the explicit time dependence of the  field
operators is derived. The normal ordering contribution then becomes:
\begin{eqnarray}
g_n(x,\tau)&=&{\rm sgn}(\tau)\prod_{\nu=\rho,\sigma}
e^{-\alpha \pi {\rm sgn}(\tau)\left[\theta_\nu(x,\tau),
\varphi_\nu(0,0)\right]/2}\nonumber\\
&=& {\rm sgn}(\tau)e^{{i\alpha\pi x {\rm sgn}(\tau)}/L}
{}~,~~~~x\neq0~,\label{compact normal}\end{eqnarray}
where the explicit form of the commutator is given in
Eq.~(\ref{commutation rho},\ref{commutation sigma}).
For later use we also give the limiting result
for $\tau\rightarrow0^{\pm}$,
\begin{equation}
g_n(x,0^{\pm})=\pm\biggl({1+\bar{z}(x)\over
1+{z}(x)}\biggr)^{\pm\alpha/2}~,
\label{gzero}
\end{equation}
where $z(x)=\exp(2\pi ix/L-\epsilon)$ ($\epsilon L$ is a short--
distance cutoff of the order of the lattice constant).
Eq. (\ref{gzero}) is valid for arbitrary $x$.

Next we compute the expectation value in Eq.~(\ref{Greens Boson}).
Going to the Fourier representation,
\begin{equation}
\Pi_\nu({\bf x})={1\over \beta L}
\sum_{\bf k}e^{i{\bf k}\cdot{\bf x}}\Pi_\nu({\bf k})~,
\end{equation}
we perform the integral over the canonical momenta
$\Pi_\rho$, $\Pi_\sigma$ with the result:
\begin{eqnarray}
<\exp\biggl[&-&is_\nu \sqrt{\pi\over 2}\biggl(\int^xdx^\prime
\Pi_\nu(x^\prime,\tau)\nonumber\\&-&
\int^0dx^\prime\Pi_\nu(x^\prime,0)\biggr)\biggr]>
\nonumber\\=~\exp\Biggl[&-&{\pi\over 4\beta L u_\nu
K_\nu}\sum_{k,\omega}
{|e^{i{\bf k}\cdot{\bf x}}-1|^2\over k^2}\Biggr]~.
\label{Pi integral}
\end{eqnarray}
Note that the exponent on the r.h.s. of Eq.  (\ref{Pi integral})
contains a negative infinite constant due to the summation
over Matsubara frequencies, however,
as we shall see below, this contribution
is exactly cancelled when integrating out the field
$\varphi_\nu$.

The functional integral over $\varphi_\nu$ gives the contribution:
\begin{eqnarray}
I_\nu ({\bf x})=
<\exp\biggl[&\alpha& s_\nu i\sqrt{\pi\over
2}(\varphi_\nu(x,\tau)-\varphi_\nu(0,0))
\nonumber\\&+&\sqrt{\pi\over 2}{1\over u_\nu
K_\nu}\biggl(\int^xdx^\prime\partial_\tau
\varphi_\nu(x^\prime,\tau)\nonumber\\
&-&\int^0dx^\prime\partial_\tau
\varphi_\nu(x^\prime,0)\biggr)\biggr]>~,
\label{varphi integral}\end{eqnarray}
where the last term in the exponential comes from the shift in
the conjugate momentum as discussed in Sec. \ref{Lagrangian
section}.

The propagator for the spin field $\varphi_\sigma$ is given by
\begin{equation}
D_\sigma({\bf k})=K_\sigma^{-1}(u_{\sigma}^{-1}
\omega^2+u_\sigma k^2)~.
\label{sigma propagator}
\end{equation}
Using this result and Eq.~(\ref{propagator})
we obtain for $I_\nu$
\begin{eqnarray}
I_\nu ({\bf x})= \exp
\Biggl[&&{\pi\over 4 \beta L}\sum_{\bf k}D_\nu^{-1}
({\bf k})\nonumber\\
&&\times\biggl(|e^{i{\bf k}\cdot{\bf x}}-1|^2\biggl({\omega^2 \over
k^2u_\nu^2K_\nu^2}-1\biggr)\nonumber\\ &&
+{4i\alpha s_\nu \omega\over k u_\nu K_\nu}
cos({\bf k}\cdot{\bf x}) \biggr)~\Biggr]~.
\label{intermediate varphi}
\end{eqnarray}
The first term in the exponent of Eq.  (\ref{intermediate varphi})
diverges due to the sum over Matsubara frequencies. This term is
positive, and the divergent part is cancelled by the contribution of
Eq. (\ref{Pi integral}). To see this
it is useful to write:
\begin{equation}
{\omega^2\over k^2D_\nu({\bf k})}=
{{\omega^2- u_\nu K_\nu D_\nu ({\bf k})}\over k^2D_\nu({\bf k})}
+{u_\nu K_\nu\over k^2}~.
\label{isolate}\end{equation}
When inserted in Eq. (\ref{intermediate varphi}),
the second term on the right hand side cancels exactly the
contribution of Eq. (\ref{Pi integral}).

The thermodynamic average in Eq. (\ref{Greens Boson}) then
becomes:
\begin{eqnarray}
<T\psi_{s\alpha} ({\bf x})\psi_{s\alpha}^{\dagger} (0)>={1\over
\epsilon L}
\prod_{\nu=\rho,\sigma}
\exp\Biggl[&&-{\pi\over 4\beta L}\sum_{\bf k}D_\nu^{-1}({\bf
k})\nonumber\\
&&~~\biggl(|e^{i {\bf k}\cdot{\bf x}}-1|^2\biggl(1+
{u_\nu K_\nu D_\nu({\bf k})-\omega^2\over
u_\nu^2 K_\nu^2 k^2}\biggr)\nonumber\\
&&~~~+{4i\alpha\omega s_\nu \over u_\nu
K_\nu k}\sin(kx)\sin(\omega \tau)\biggr)~\Biggr]\nonumber\\
\times g_n({\bf x})&&~.
\label{Greens result}
\end{eqnarray}
Note that the exponent in Eq. (\ref{Greens result}) has a real and
imaginary part coming from both the thermodynamic average and
the normal
ordering procedure. This implies that the Green function has an
oscillatory component and decays with a power law as expected.
The frequency  and momentum sums in Eq. (\ref{Greens result})
are performed in Appendix B.  The result is:
\begin{mathletters}
\begin{eqnarray}
<T\psi_{s\alpha} ({\bf x})\psi_{s\alpha}^{\dagger} (0)>={1\over
\epsilon L}
\exp\Biggl[&-&{1\over 4}\sum_{\beta=\pm}({C_\beta \over K_\rho}
+{K_\rho u_\rho \over v_\beta}F_\beta)
\ln[|1-z(x+iv_\beta|\tau|)|/\epsilon]\nonumber\\
&-&{1\over 4}(K_\sigma+{1\over K_\sigma})
\ln[|1-z(x+iu_\sigma|\tau|)|/\epsilon]\nonumber\\
&-&{\rm sgn}(\tau){\alpha s\over 4}
\ln{1-\bar{z}(x+iu_\sigma |\tau|)\over 1-z(x+iu_\sigma
|\tau|)}\nonumber\\
&-&{\rm sgn}(\tau){\alpha\over 4}\sum_{\beta=\pm}F_\beta
\ln{1-\bar{z}(x+iv_\beta|\tau|)\over 1-z(x+iv_\beta
|\tau|)}\Biggr]\nonumber\\\times&&g_n({\bf x})~.
\label{Greens 2}
\end{eqnarray}\end{mathletters}
We have introduced
effective velocities induced by the phonon coupling:
\begin{equation}
v_\pm^2={1\over 2}(u_\rho^2+c^2)\mp{1\over 2}\sqrt{
(u_\rho^2-c^2)^2+4b^2}~,
\label{velocities}
\end{equation}
as well as the abbreviations:
\begin{mathletters}
\begin{eqnarray}
C_\pm&=&{u_\rho\over v_\pm}{v_\pm^2 -c^2+b^2/u_\rho^2\over
v_\pm^2-v_\mp^2}
\label{C plus minus}\\
F_\pm&=&{v_\pm^2-c^2\over v_\pm^2-v_\mp^2}~.
\label{F plus minus}\end{eqnarray}\end{mathletters}

The thermodynamic limit corresponds to taking
$|x+iv\tau|/L<<1$, which in turn implies
$1-z(y)\rightarrow -i 2\pi y/L+\epsilon$.
We can now give the final expression for the Green
function  in the thermodynamic limit  using
Eq. (\ref{Greens 2}) and Eq.~(\ref{Greens}):
\begin{eqnarray}
G_s(x,\tau)&=&{1\over\epsilon L}
\left|{\epsilon L/2\pi\over x+iu_\sigma\tau}\right|^{K_\sigma/4
+1/4K_\sigma}
\nonumber\\
&~&\times\prod_{\beta=\pm}\left|{\epsilon L/2\pi\over
x+iv_\beta\tau}\right|^{K_\rho
u_\rho F_\beta/4v_\beta+C_\beta/4K_\rho}\nonumber\\
&~&\times\sum_{\alpha=\pm}e^{i \alpha k_F x}
\biggl({s \alpha  x+iu_\sigma |\tau|\over
i|x+iu_\sigma \tau|}\biggr)^{sgn(\tau)/2}\nonumber\\
&~&\times\prod_{\gamma=\pm}
\biggl({ \alpha   x+iv_\gamma |\tau|\over
i|x+iv_\gamma \tau|}\biggr)^{sgn(\tau)F_\gamma /2}
g_n(x,\tau)\nonumber\\
{}~.
\label{end result Greens}
\end{eqnarray}

Near ${\bf x}=0$ one has to add $\epsilon L/2\pi$ to $v|\tau|$.
At large distances and/or times $g_n$ becomes unity and
the power law decay of the
Green function becomes then (in obvious notation)
\begin{equation}
G_s({\bf x})\propto |x,\tau|^{-1-\delta}~,
\label{power law}
\end{equation}
with
\begin{equation}
\delta={K_\sigma\over 4}
+{1\over 4 K_\sigma}-1
+{B\over 4K_\rho}+{A K_\rho\over 4}~,
\label{delta}\end{equation}
and the electron--phonon parameters A and B are defined by
\begin{mathletters}
\begin{eqnarray}
A(u_\rho,c,b)&\equiv&\sum_{\beta=\pm}{u_\rho\over
v_\beta}F_\beta
={u_\rho\over v_++v_-}\biggl(1+{c^2\over v_+v_-}\biggr)\nonumber
\\ \label{A}\\
B(u_\rho,c,b)&\equiv&\sum_{\beta=\pm}C_\beta={u_\rho\over
v_++v_-}
\biggl(1+{v_+v_-\over u_\rho^2}\biggr)\label{B}
\end{eqnarray}\end{mathletters}
These exponents play an important role in the following discussion
of the ordering fluctuations. For later use
we only note here
their limiting behavior. Defining a critical charge velocity by
the equality $b/u_\rho c=1$, or alternatively:
\begin{equation}
{u_\rho^*\over K_\rho^*}={g^2\over\hbar\zeta c^2}~,
\label{critical}\end{equation}
we see from Eq.~(\ref{velocities})
that $v_+^2$ tends to zero, and $v_-^2$ to ${u_\rho^*}^2+c^2$, as
$u_\rho$ approaches this critical velocity $u_\rho^*$ from above.
As a consequence, the exponent $A$ diverges to infinity and $B$
decreases
to the finite value $u_\rho^*/\sqrt{{u_\rho^*}^2+c^2}$ as
$u_{\rho}\rightarrow
u_\rho^*$.
For $u_\rho < u_\rho^*$ the velocities become
complex and the model becomes unphysical.
Thus we must require
that $u_\rho/K_\rho\ge u_\rho^*/K_\rho^*$, or equivalently that
$b/(cu_\rho)\leq 1$, the equality sign defines the Wentzel--Bardeen
singularity\cite{footnote0}.
We emphasize that this
singular
behavior is a non-perturbative effect of the electron-phonon
coupling
and originates from the instability of the effective electron
propagator
as pointed out after Eq.~(\ref{propagator}).
We also note  here that the limits $g\rightarrow 0$ and
$u_\rho\rightarrow u_\rho^*$ do not commute.

In the limit of vanishing
phonon coupling constant, the charge and spin propagator have
an identical structure, and $A,B \rightarrow 1$.
The result is:
\begin{equation}
\lim_{g\rightarrow 0}G_s({\bf x})\propto |x,\tau|^{-K_\rho/4-1/
4K_\rho
-K_\sigma/4-1/ 4K_\sigma}~.
\label{zero coupling}
\end{equation}
In particular, for non--interacting electrons $K_\rho=K_\sigma=1$,
we
recover the $1/x$
power law dependence of the Green function of the free electron
gas.

To illustrate our results with a specific model of interacting electrons,
we plot the quantity $\delta$ for the Hubbard model in Fig.
\ref{fig1}a
and \ref{fig1}b. The Luttinger liquid parameters $K_\rho$, $u_\rho$
are determined numerically later on in Sec. \ref{Hubbard section}.

We first plot $\delta$ for a quarter filled band as a
function of $U$ in Fig. \ref{fig1}a, for several values of
the phonon coupling parameter. For small phonon coupling, $\delta$
increases monotonically with increasing $U$. As the phonon coupling
is further increased, $\delta$ acquires a minimum.

Next, we choose $U=2$ and vary the filling factor from zero to half
filling in Fig. \ref{fig1}b. $\delta$ increases dramatically when these
two extremes are reached, as the correlation effects between
electrons
dominate the physics in both cases. In particular,
$\delta\rightarrow\infty$ at the Wentzel--Bardeen singularity.

\subsection{Momentum distribution function}
\label{momentum subsection}

We now examine the effect of the phonon coupling on
the momentum distribution function which is given by
\begin{equation}
N_\alpha(k)\equiv {1\over 2}\sum_s
\int_{-\infty}^{+\infty}d x~e^{i (k-\alpha k_F)x}
<\psi_{s\alpha}^\dagger(x,0)
\psi_{s\alpha}(0)>~.
\label{momentum}\end{equation}
Next, we insert the result of Eqs.  (\ref{power law}) and (\ref{delta})
at $\tau=0^-$\cite{footnote1}, use that in the thermodynamic limit
$g_n(x,0^{\pm})=\pm i\alpha|x+i\alpha|/(x+i\alpha)$ (see
Eq.~(\ref{gzero})),
and find (setting $a_o=\epsilon L/2\pi$),
\begin{equation}
N_\alpha(k)=-\alpha\int_{-\infty}^{+\infty}{d x\over 2 \pi i}~
{e^{i (k-\alpha k_F) x}\over|x+ia_o|^{\delta}}{a_o^\delta\over
x+i\alpha a_o}~.
\label{integral}
\end{equation}
As pointed out in Ref. \cite{Mahan}, for a system with
$\delta>0$, the exponential in the integrand of Eq.  (\ref{integral}) is
not necessary any more to insure the convergence of the integral. For
$k\simeq \alpha k_F$, we may omit this oscillatory factor,
and obtain the distribution function evaluated at $k=\alpha k_F$
\begin{eqnarray}
N_\alpha(\alpha k_F)&=&\int_0^{\infty}
{dy\over\pi}
{}~ (1+y^2)^{- (1+\delta/2}~, \nonumber\\
&=&{\Gamma (1/2+\delta/2) \over 2\sqrt{\pi}\Gamma
(1+\delta/2)}~,
\label{distribution}
\end{eqnarray}
where $\Gamma(x)$ denotes the Gamma function. In the limit
$\delta\rightarrow 0$,
$N_\alpha(\alpha k_F)=1/2$, which differs from the Fermi step
function obtained for the free system at $\delta=0$.
The momentum distribution function has a
finite
value at the Fermi wave vector, but its derivatives are all singular at
that point (see below). This result was previously
found in Ref. \cite{Mahan} for a system of interacting fermions but
without
phonons.
The fact that the momentum distribution function has no jump
at the Fermi surface for interacting fermions
expresses the failure of the
Fermi liquid quasiparticle
picture to capture the low temperature behavior of interacting
one--dimensional systems, which has been known for the Luttinger
model since the
pioneering work of Mattis and Lieb \cite{Mattis}.

We can further determine the behavior  of the momentum
distribution
function around $k=k_F$
by noting that
\begin{equation}
N_{+}(k)={1\over \pi} \int_0^\infty
dx {\cos{(a_ox(k-k_F))}\over
(x^2+1)^{1+\delta/2}}\cdot(1+{\cal O}(\sqrt{|k-k_F|}))~.
\label{near Fermi1}\end{equation}
Evaluating the tabulated integral we then find
\begin{eqnarray}
N(k\simeq k_F)
\simeq N_{+}(k_F)-c{\rm sgn}(k-k_F)|k-k_F|^{\delta}~,\nonumber
\\
\label{near Fermi2}\end{eqnarray}
with $c=a_o^{\delta-1/2}\Gamma(-1-\delta)/\Gamma(-\delta/2)$.
Eq. (\ref{near Fermi2}) exhibits the singular dependence
of the momentum distribution function at $k=k_F$.
Note that for the pure electron system ($g\rightarrow 0$),
taking $K_\sigma=1$, we recover the exponent of the momentum
distribution function given in Ref. \cite{Schulz}, namely:
\begin{equation}
\delta\rightarrow {K_\rho\over 4}+{1\over 4K_\rho}-{1\over 2}~.
\label{momentum exponent zero}\end{equation}

\subsection{Free electron gas coupled to phonons}
\label{free subsection}

In Eq. (\ref{delta}), both the effect of the interaction between
electrons
and the effect of the coupling with the phonons are included. In this
section,
we specialize to the case of free electrons coupled to phonons in
order to
clarify the role played by the phonons.
In this case, $\delta$ takes the form:
\begin{equation}
\delta=A/4+B/4-1/2~.\label{delta 0}\end{equation}
It is convenient to rewrite $A$ in the following way:
\begin{equation}
A=B+{b^2\over u_\rho v_+v_-(v_++v_-)}~.
\end{equation}
Noting that
\begin{equation}
B^2=1-{b^2/[u_\rho(v_++v_-)]^2}\leq 1,
\label{Bnew}\end{equation}
we calculate
\begin{equation}
(A+B)^2=4+\biggl({b\over v_+v_-(v_++v_-)}\biggr)^2\biggl({b^2\over
u_\rho^2}+4v_+v_-\biggr)~.\end{equation}
With Eq. (\ref{delta 0}), and $A,B\ge 0$ this implies that $\delta>0$,
for $g\neq0$.
If we start from a free Fermi gas,
and couple it to phonons, the system
is driven away from the free Fermi behavior and the Fermi
distribution
function is totally destroyed for an arbitrarily small
phonon coupling strength $g$.
\section{Ordering fluctuations in one--dimension}
\label{ordering section}

In this section, we determine which type of fluctuations dominate
in an interacting electron gas coupled to phonons.

\subsection{Correlation functions}
\label{correlation subsection}

The
tendency
of the system towards ordering
manifests itself in divergent correlation functions,
which describe the response of the system to external perturbations.
Thus, to characterize
the type of long-range order (collective state), which might be
realized in
higher dimensions,
we follow standard practice \cite{Emery,Solyom,Schulz} and consider
response functions of the type
\begin{equation}
R(x,t)=-i\Theta (t)<\bigl[O^\dagger(x,t),O\bigr]>,
\end{equation}
where the charge-density wave (CDW) response is generated by the
operator,
$O_{CDW}=[\psi_{+\uparrow}^\dagger \psi_{-\uparrow}+
\psi_{+\downarrow}^\dagger \psi_{-\downarrow}]/2$,
the transverse spin-density wave (SDW) response by
$O_{SDW}=\psi_{+s}^\dagger \psi_{-,-s}$,
singlet pairing (SS) by $O_{SS}=\psi_{+s}^\dagger \psi_{-,-s}^\dagger$,
and triplet pairing (TS) by $O_{TS}=\psi_{+s}^\dagger \psi_{-
,s}^\dagger$.
The corresponding two-particle Green functions in Matsubara
representation
read then:
\begin{mathletters}
\begin{eqnarray}
N(x,\tau)&=&-ie^{i2k_Fx}<T\psi_{-\uparrow}^\dagger
(x,\tau)\psi_{+\uparrow}
(x,\tau)\psi_{+\uparrow}^\dagger \psi_{-\uparrow}>
\nonumber\\ \label{c d w}\\
\chi(x,\tau)&=&-ie^{i2k_Fx}<T\psi_{-\downarrow}^\dagger(x,\tau)
\psi_{+\uparrow}(x,\tau)\psi_{+\uparrow}^\dagger
\psi_{-\downarrow}>\nonumber\\ \label{s d w}\\
\Delta_s(x,\tau)&=&-i<T\psi_{-\downarrow}(x,\tau)
\psi_{+\uparrow}(x,\tau)\psi_{+\uparrow}^\dagger
\psi_{-\downarrow}^\dagger>\label{singlet}\\
\Delta_t(x,\tau)&=&-i<T\psi_{-\uparrow}(x,\tau)
\psi_{+\uparrow}(x,\tau)\psi_{+\uparrow}^\dagger
\psi_{-\uparrow}^\dagger>~,\label{triplet}
\end{eqnarray}
\end{mathletters}
with the notation $\psi_{\pm s}$ defined in Eq.
(\ref{decomposition}).
$N$ represents the correlation function associated with the
CDW state, which for example occurs when the nearest
neighbor
repulsion is large compared to the other parameters of the model.
$\chi$ represents the transverse SDW response function, which
describes the onset of antiferromagnetic ordering.
$\Delta_s$ ($\Delta_t$)
gives the probability amplitude for a singlet (triplet)
Cooper pair emitted at the origin of space and time
to be annihilated at $x$ and $\tau$, and thus characterizes
the superconducting fluctuations. To simplify notation we have
retained
only a representative term in the correlation functions, and this term
will determine the correct small frequency/momentum behavior of
the whole
correlation function.

Using the bosonized version of $\psi_{\pm s}$ in conjunction with
the
decomposition into charge and spin of Eqs. (\ref{charge}) and
(\ref{spin}), we obtain (for $\tau>0$):
\begin{mathletters}
\begin{eqnarray}
N(x,\tau)=-{ie^{i2k_Fx}\over (\epsilon L)^{2}}
&<&\prod_{\nu=\rho,\sigma}
\exp\biggl[i\sqrt{2\pi}(\varphi_\nu(x,\tau)-\varphi_\nu(0,0))
\biggr]>
\label{c d w boson}\\
\chi(x,\tau)=-{ie^{i2k_Fx}\over (\epsilon L)^{2}}
&<&\exp\biggl[-i\sqrt{2\pi}\biggl(\int^x
dx^\prime\Pi_\sigma(x^\prime,\tau)-\int^0
dx^\prime\Pi_\sigma(x^\prime,0)\biggr)\biggr]
\nonumber\\
&~&exp\biggl[i\sqrt{2\pi}
(\varphi_\rho(x,\tau)-\varphi_\rho(0,0))
\biggr]>\label{s d w boson}\\
\Delta_s(x,\tau)={-i\over (\epsilon L)^{2}}
&<&\exp\biggl[-i\sqrt{2\pi}\biggl(\int^x
dx^\prime\Pi_\rho(x^\prime,\tau)-\int^0
dx^\prime\Pi_\rho(x^\prime,0)\biggr)\biggr]\nonumber\\
&~&\exp\biggl[i\sqrt{2\pi}(\varphi_\sigma(x,\tau)-
\varphi_\sigma(0,0))
\biggr]>\label{singlet boson}\\
\Delta_t(x,\tau)={-i\over (\epsilon L)^{2}}
&<&\exp\biggl[-i\sqrt{2\pi}\biggl(\int^x
dx^\prime\Pi_\rho(x^\prime,\tau)
-\int^0 dx^\prime\Pi_\rho(x^\prime,0)\biggr)\biggr]
\nonumber\\&~&\exp\biggl[-i\sqrt{2\pi}\biggl(\int^x
dx^\prime\Pi_\sigma(x^\prime,\tau)-\int^0
dx^\prime\Pi_\sigma(x^\prime,0)\biggr)\biggr]>~,\nonumber\\
\label{triplet boson}
\end{eqnarray}
\end{mathletters}

The functional integrations which are implied in Eqs. (\ref{c d w
boson})
to (\ref{triplet boson}) have already been encountered in the
calculation
of the single-particle Green function in Sec. \ref{single section}. We
therefore omit the details of this calculation. Also, because we are
only
interested in the long distance and long time properties of the
correlation functions, we have ignored the oscillatory terms arising
from
the normal ordering procedure.
The results are:
\begin{mathletters}
\begin{eqnarray} N(x,\tau)&=&-{ie^{i2k_Fx}\over (\epsilon L)^{2}}
\exp \Biggl[-K_\rho\sum_{\beta=\pm}
{u_\rho \over v_\beta}F_\beta \ln{|1-z(x+iv_\beta\tau)|/\epsilon}
\nonumber\\
&~&~~~~~~~~~~-K_\sigma\ln{|1-z(x+iv_\sigma \tau)|/\epsilon}
\Biggr]\label{result c d w}\\
\chi(x,\tau)&=&-{ie^{i2k_Fx}\over (\epsilon L)^{2}}
\exp\Biggl[-K_\rho\sum_{\beta=\pm}
{u_\rho \over v_\beta}F_\beta
\ln{|1-z(x+iv_\beta \tau)|/\epsilon}\nonumber\\
&~&~~~~~~~~~~-{1\over K_\sigma}\ln{|1-z(x+iu_\sigma
\tau)|/\epsilon}
\Biggr]\label{result s d w}\\
\Delta_s(x,\tau)&=&{-i\over (\epsilon L)^{2}}
\exp\Biggl[-{1\over
K_\rho}\sum_{\beta=\pm}C_\beta \ln{|1-z(x+iv_\beta \tau)|/
\epsilon}\nonumber\\
&~&~~~~~~~~~~-K_\sigma\ln{|1-z(x+iu_\sigma \tau)|/\epsilon}\Biggr]
\label{result singlet}\\
\Delta_t(x,\tau)&=&{-i\over (\epsilon L)^{2}}
\exp\Biggl[-{1\over K_\rho}
\sum_{\beta=\pm}C_\beta\ln{|1-z(x+iv_\beta
\tau)|/\epsilon}\nonumber\\
&~&~~~~~~~~~~-{1\over K_\sigma}\ln{|1-z(x+iu_\sigma
\tau)|/\epsilon}\Biggr],
\label{result triplet}
\end{eqnarray}
\end{mathletters}
with $C_\pm$ and $F_\pm$ defined in Eqs. (\ref{C plus minus}) and
(\ref{F plus minus}).

The thermodynamic limit is then taken to obtain the power law
behavior of the correlation functions:
\begin{mathletters}
\begin{eqnarray}
N(x,\tau)&\propto&e^{i2k_Fx}\prod_{\beta=\pm}|x+iv_\beta\tau|^{-
u_\rho K_\rho
F_\beta/v_\beta}|x+iu_\sigma \tau|^{-K_\sigma}
\nonumber\\
\label{exponent N}\\
\chi(x,\tau)&\propto&e^{i2k_Fx}\prod_{\beta=\pm}|x+iv_\beta\tau|^{
-u_\rho K_\rho
F_\beta/v_\beta}|x+iu_\sigma \tau|^{-1/K_\sigma}
\nonumber\\ \label{exponent chi}\\
\Delta_s(x,\tau)&\propto&\prod_{\beta=\pm}|x+iv_\beta\tau|^{-
C_\beta/
K_\rho} |x+iu_\sigma \tau|^{-K_\sigma}
\label{exponent singlet}\\
\Delta_t(x,\tau)&\propto&\prod_{\beta=\pm}|x+iv_\beta\tau|^{-
C_\beta/
K_\rho} |x+iu_\sigma \tau|^{-1/K_\sigma}~.
\label{exponent triplet}\end{eqnarray}
\end{mathletters}

\subsection{Divergence criteria}
\label{divergence subsection}

The indication that a given type of flucuations dominates
in the system is provided by the divergence of the Fourier transform
of the corresponding correlation function at low frequencies and
small momentum relative to $q=2k_F$ ($q=0$) for $N$ and $\chi$
($\Delta_s$ and
$\Delta_t$).

Inspecting Eqs. (\ref{exponent N}) to (\ref{exponent triplet}), we
notice
that such a divergence can only originate from the power law
behavior
of the correlation functions at large distances and imaginary times.
Whether or not there is a divergence then depends on the exponents
which were calculated in the previous section.
In particular, we note that
\begin{mathletters}\begin{equation}
|x+iv_{max}\tau|^{-\gamma_k}\leq |C_k (x,\tau)|
\leq |x+iv_{min}\tau|^{-\gamma_k},
\end{equation}\end{mathletters}
where $C_k$ is given by the corresponding rhs of Eqs.(\ref{exponent
N}) to
(\ref{exponent triplet}), $\gamma_k$ by  the associated sum of
exponents,
and $v_{max/min}=max/min\{v_+,v_{-},u_\sigma\}$. From this, we
obtain
immediately the divergence criteria for the Fourier transforms of the
correlation functions.
The condition for CDW fluctuations to occur is then,
from Eq. (\ref{exponent N}):
\begin{equation}
A K_\rho +K_\sigma\leq 2~.~~~~~~{\rm (CDW)~~~~}
\label{criterion
C D W}\end{equation}
{}From Eq. (\ref{exponent chi}), SDW fluctuations occur when
\begin{equation}
A K_\rho+{1\over K_\sigma}\leq 2~.~~~~~~~{\rm
(SDW)~~~~}
\label{criterion S D W}\end{equation}
{}From Eq. (\ref{exponent singlet}), singlet Cooper pairs fluctuations
are present in the system when:
\begin{equation}{B\over K_\rho}+K_\sigma\leq 2~.~~~~~~~{\rm
(singlet)}
\label{criterion singlet}\end{equation}
Finally, we will have triplet superconducting fluctuations when:
\begin{equation}
{B\over K_\rho}+{1\over K_\sigma}\leq 2~.~~~~~~{\rm
(triplet)}\label{criterion triplet}
\end{equation}
For models with interactions which do not depend on the
spin ($SU(2)$ symmetry), such as the Hubbard model,
$K_\sigma=1$, and the criteria for CDW
and for SDW order (for singlet and for triplet pairing) are
a priori the {\it
same}.
However we expect SDW and triplet Cooper pairs to dominate:
CDW order will not occur because
SDW order builds up before the longer range type of
fluctuations have time to settle in the system \cite{Emery}.
Similarly, triplet superconductivity will dominate over singlet
superconductivity. This is not noticeable in the criteria of
Eqs. (\ref{criterion C D W}) to (\ref{criterion triplet}) because
additional logarithmic corrections
have been neglected here
\cite{Logarithm}. For the electronic models
we are considering in this paper,
these corrections arise from marginally irrelevant
operators in the spin channel, which effectively generate longer tails for the
SDW (triplet) correlation function than for the CDW (singlet)
correlation functions. These operators have been omitted in
Eq. (\ref{nu Hamiltonian}) for simplicity.
If both divergence criteria for spin waves and triplet Cooper pairs
are met for a given set of parameters, the correlation function
``which diverges the most'' determines the phase of the system. Note
that the criteria of Eqs. (\ref{criterion C D W})--(\ref{criterion
triplet})
are valid for {\it any} interacting electron system with a
finite range interaction: what remains to be done is to study the
dependence
of $K_\rho$ and $u_\rho$ on the interaction parameter(s) of the
electron
Hamiltonian and the
filling factor for the electrons. Note that for repulsive interactions,
we have $K_\rho<1$.

At $K_\rho=K_\sigma=1$, the criteria for SDW order
and triplet pairing take the simple form:
\begin{mathletters}
\begin{eqnarray}
A\leq1~&.&~~~~~~~~{\rm (SDW)}
\label{non criterion S D W}\\
B\leq1~&.&~~~~~~~~{\rm (Cooper~pairs)}
\label{non criterion triplet}\end{eqnarray}\end{mathletters}
from the results of Sec. \ref{free subsection} and noting that
\begin{equation}
A^2={1-b^2/[c(v_++v_-)]^2\over 1-b^2/(c u_\rho)^2}~,\label{A
square}\end{equation} with $u_\rho<v_++v_-$  and $c>0$, we
conclude that $A>1$,
and Eq. (\ref{non criterion S D W}) is never satisfied, while Eq.
(\ref{non criterion triplet})
is always satisfied according to Eq.~(\ref{Bnew}).
The Cooper instability is
always present for a system
of non-interacting electrons, regardless of the magnitude and sign of
the phonon coupling constant.

When no phonons are present in the system, $A=B=1$, and a system
of electrons with repulsive on--site interaction always has
$K_\rho<1$,
which
implies from the inequality (\ref{criterion S D W}) that SDW (CDW)
fluctuations
dominate over superconducting fluctuations, as expected.

A check of the above results can be obtained
by considering
non--interacting electrons ($K_\rho=K_\sigma=1$) coupled to
phonons
in the limit where the speed of sound is much larger than the Fermi
velocity.
 If we set $c\rightarrow\infty$,
keeping $b/cu_\rho$ finite ($0<b/cu_\rho<1$), the propagator
of Eq.  (\ref{propagator}) becomes:
\begin{equation}
D_\rho(k)\simeq (K_\rho u_\rho)^{-1}(\omega^2+u_\rho^2k^2-
b^2k^2/c^2
+c^{-2}O(\omega^2k^2))~.\label{infinite c propagator}\end{equation}
This propagator now corresponds to a system of electrons (with no
phonons)
with an
{\it attractive} interaction without retardation effects, as
in the original Cooper problem.
We should therefore be able to reproduce the result of Luther
and Peschel for spinless fermions \cite{Luther},
which predicts strong pairing fluctuations
for the attractive case.
In the limit $c\rightarrow \infty$, we obtain from Eqs.
 (\ref{A}) and  (\ref{B}) the result that $A= (1-(b/cu_\rho)^2)^{-1/2}$
and
$B= (1-(b/cu_\rho)^2)^{1/2}$.
The pairing correlation function for triplet pairing diverges,
according to Eq. (\ref{criterion triplet}),
but the criterion for SDW order is not met.
The attractive
interaction induces strong
pairing fluctuations, in anology with the Cooper problem and the
result of
Luther and Peschel for spinless fermions.

\section{Hubbard model}
\label{Hubbard section}

In this section, we shall consider the effect of phonons
on the correlation function exponents derived in the previous
sections,
for specific interacting electron models. Most of the discussion will
consider a Hubbard model with on--site repulsion parameter $U$:
\begin{equation}
H_{el}=-t\sum_{i,s} (c^\dagger_{is}c_{i+1,s}+c^\dagger_{i+1,s}c_{is})
+U\sum_in_{i\uparrow}n_{i\downarrow}~,
\label{Hubbard model}
\end{equation}
with $t$ the hopping matrix element.
We shall also consider the extended Hubbard model with nearest
neighbor
repulsion parameter $V$ in the limit $U\rightarrow\infty$, where
analytical expressions for $u_\rho$ and $K_\rho$ can be found at
quarter
filling.

\subsection{Weak interactions}
\label{weak subsection}

We first consider the on--site parameter as a perturbation.
The charge velocity is then given by the Fermi velocity:
\begin{equation}
u_\rho=v_F=2t \sin(n\pi/2)~,\label{charge zero}\end{equation}
for a system which has
$n$ electrons per site (half filling corresponds to $n=1$).
The shift in energy (per site) due to the on--site repulsion term
in Eq.(\ref{Hubbard model}) is then
$\Delta E=Un^2/4$. One therefore deduces the first order expression
for $K_\rho$:
\begin{equation}
K_\rho=1-U/\pi v_F+O(U^2)~.\label{K perturbation}
\end{equation}
Inserting the latter expression in the criteria for SDW order
and triplet pairing, we obtain:
\begin{mathletters}\begin{eqnarray}
{2U\over\pi v_F}&>&{b^2(c^2+2v_+v_-)\over v_+^2v_-^2(v_++v_-)^2}
{}~~~~~({\rm SDW})\label{weak S D W}\\
{2U\over\pi v_F}&<&{b^2\over v_F^2(v_++v_-)^2}~,~~~~~({\rm
Cooper~pairs})
\label{weak triplet}\end{eqnarray}
\end{mathletters}
with $v_\pm$ given by Eq. (\ref{velocities}).
Eqs. (\ref{weak S D W})
and (\ref{weak triplet}) displays
the effects mentioned in the Introduction:
the coupling to phonons will induce
Cooper pairs if the coupling constant is
large enough to overcome the instantaneous
repulsion between the electrons. Similarly,
in order to have SDW fluctuations, the
on--site repulsion parameter must be
large enough to overcome the retarded
electron atttractive interaction
mediated by the phonons. For sufficiently
small phonon coupling $v_+\rightarrow c$ and
$v_-\rightarrow v_F$, the above inequalities are mutually
excusive: there is a region
in parameter space:
\begin{equation}
{b^2\over v_F^2(v_F+c)^2}<{2U\over\pi v_F}<{b^2(1+2v_F/c)\over
v_F^2(v_F+c)^2}
\label{region}\end{equation}
which separates the SDW order and the triplet pairing phase.

Note that the perturbation expansion breaks down in both limits
$n\rightarrow 0$ and $n\rightarrow 1$. In these two cases,
we shall see in Sec. \ref{arbitrary subsection} that
$K_\rho\rightarrow
0.5$, which signals that
the correlation effects are important. Near $n=1$, Umklapp scattering
becomes relevant, and the system feels the proximity of the
metal-insulator transition.

We plot the phase diagram of the system for $U=0.01$
in Fig. \ref{fig2}, using the
perturbative result of Eq. (\ref{K perturbation}), as a function
of the filling factor $n$ and the phonon coupling parameter
$b/cu_\rho$.
We choose $c\sim \epsilon L t/\hbar$, which is reasonable for
heterostructure systems where $c\simeq 5\times 10^5 [cm/s]$ and
$v_F\simeq 10^6$.
The superconducting phase is separated from the
SDW phase by a metallic region, as suggested by the
small coupling result of Eq. (\ref{region}).

\subsection{Hubbard model at arbitrary filling factor}
\label{arbitrary subsection}

To determine the Luttinger liquid parameters $K_\rho$ and
$u_\rho$
for arbitrary $U$ and arbitrary filling factor $n$, we follow the
method
proposed by Ref. \cite{Schulz}. We calculate the ground state
energy by solving numerically the integral equation of Lieb and Wu
\cite{Lieb,Shiba}:
\begin{equation}
f(k)={1\over 2\pi}+{4\over u}\cos(k)\int_{-Q}^Q dk^\prime
R\biggl({4\over u}(\sin(k)-\sin(k^\prime))\biggr)f(k^\prime)~,
\label{integral equation 1}\end{equation}
where $u=U/t$, and $f(k)$ is a distribution function evaluated at
pseudo--momentum $k$. $k$ take values in the interval $[-Q,Q]$,
and $Q<\pi$ depends on the filling factor through the relation
$\int_{-Q}^Qdk f(k)=n$. The function R(x) is the cosine Fourier
transform
of $1/(1+e^x)$ evaluated at $x/2$.
With these notations, the ground state energy is given by:
\begin{equation}
E_G=-2t\int_{-Q}^Qdk \cos(k)f(k)~.\label{ground energy}
\end{equation}
A first relation between $K_\rho$ and $u_\rho$ is obtained by
noting
\cite{Schulz,Schulz2}:
\begin{equation}
{\partial^2 E_G\over\partial n^2}={\pi u_\rho\over 2 K_\rho}~.
\label{double derivative}\end{equation}
The velocity of the charge excitations is obtained from the particle
(or hole) excitation spectrum, which was calculated by Coll
\cite{Coll}. $u_\rho$ corresponds to the group velocity of particles
(holes) in the limit of small momentum:
\begin{equation}
u_\rho=\lim_{k\rightarrow 0}{\Delta \epsilon(k)\over\Delta p(k)}
=\biggl(2\sin(Q)-\mu\int_{-Q}^Qdk
g(k)-2t\int_{-Q}^Qdk\cos(k)g(k)\biggr)/2\pi f(Q)~,
\label{charge velocity}\end{equation}
with $\Delta\epsilon(k)$ the particle (hole) excitation energy,
$\Delta p(k)$ its associated momentum and $\mu$ the chemical
potential.
The function $g(k)$ in Eq. (\ref{charge velocity}) is the derivative of
the particle (hole) excitation distribution function in pseudo
momentum
space.
It is the solution of the integral equation:
\begin{eqnarray}
g(k)&=&-{16\over u^2}\cos(k)\cos(Q)R^\prime\biggl({4\over
u}(\sin(k)-sin(Q))\biggr)\nonumber\\
&+&~~{4\over u}\cos(k)\int_{-Q}^Q dk^\prime
R\biggl({4\over u}(\sin(k)-\sin(k^\prime))\biggr)g(k^\prime)~,
\nonumber\\
\label{integral equation 2}\end{eqnarray}
where $R^\prime(x)$ denotes the derivative of $R(x)$.

Once the parameters $u_\rho$ and $K_\rho$ are specified,
we can determine in what region of parameter space the criterion
(\ref{criterion triplet}) and (\ref{criterion S D W}) are satisfied.
The speed of sound is assumed to have the same value as in Sec.
\ref{weak subsection}.
We choose to represent the phonon coupling with the
dimensionless parameter $b/cu_\rho$. We emphasize that:
\begin{equation}
{b\over cu_\rho}\equiv {g\over c}\sqrt{K_\rho\over \hbar\zeta u_\rho}
\label{effective coupling}\end{equation}
is an {\it effective} electron--phonon parameter, which, strictly
speaking, depends of the filling factor. Nevertheless, it allows us
to display the results in an obvious fashion. First we consider the case of
$b/cu_\rho$ strictly smaller than unity (i.e. excluding the Wentzel--Bardeen
singularity).

First, we vary the filling factor $n$ and the phonon coupling
for fixed $U$.
Our results are illustrated in Figs. (\ref{fig3}a)--(\ref{fig3}c),
for $U/t=0.7$, $U/t=0.3$, $U/t=0.1$ . The regions where SDW fluctuations
dominate,
or
alternatively, superconducting fluctuations dominate, are separated
by a metallic region, as was proposed
in Sec. \ref{weak subsection}. For small phonon
coupling, SDW fluctuations
dominate for arbitrary filling factor. As the coupling is increased,
we cross the metallic region first near quarter filling, and eventually
we reach the regime of superconducting fluctuations.
For low filling factors and near half filling, correlation effects
dominate the picture, and superconducting fluctuations cannot be
sustained, even at large phonon coupling. By comparing Fig.
(\ref{fig3}a)
to Fig. (\ref{fig3}b) and Fig. (\ref{fig3}c), we observe that
the region of parameter space where superconducting fluctuations
dominate
shrinks as $U$ is increased. This demonstrates the interplay between
the
instantaneous repulsion and the phonon mediated attraction for
determining the phase of the system. We warn the reader that
due to some limitations in numerical accuracy associated with the sharp
variations of the charge velocity in the immediate neighborhood of
$n=1$, Figs. \ref{fig3}a
to \ref{fig3}c seem to indicate that the SDW phase dominates
in this limit. This is not the case, for as we shall see
in Sec. \ref{Wentzel subsection}, antiferromagnetic  fluctuations
are {\it always} suppressed for $b/cu_\rho=1$.

Next, we choose the filling factor at quarter filling
($n=1/2$), and vary $U$ and the phonon coupling, see Fig. \ref{fig4}.
Again, the metallic region separates the superconducting phase
from the
SDW
phase. At low $U$ ($U/t<0.1$), $K_\rho\approx 1$, and the Cooper
instability
pushes the system in the superconducting phase. As $U$ is further
increased,
the phase boundaries appear, and the phonon--mediated attractive
interaction is quickly overcome by the instantaneous repulsion between
electrons.
Beyond $U/t=2$, the SDW fluctuations dominate for all values of the
phonon
coupling.

\subsection{The Wentzel--Bardeen singularity}
\label{Wentzel subsection}

In the preceding section, we plotted the phase diagram
of the system as a function of the filling factor and the on site
repulsion versus the phonon coupling parameter
$b/u_\rho c$. The maximal value for this latter parameter
in Figs. \ref{fig3}
and \ref{fig4} is $1$, and corresponds to the Wentzel--Bardeen
singularity where the compressibility of
the system becomes negative. We now show
that for the Hubbard model, this
singular point is accessible by varying the filling factor
and discuss the physics in this region of parameter space.

We first reexpress the phonon coupling parameter in terms of the bare
phonon coupling constant, the charge velocity, and the
parameter $K_\rho$, as is apparent in Eq. (\ref{effective coupling}).
Next, we plot the ratio $u_\rho /K_\rho$ in Fig. \ref{fig5}
for several values of $U$, as a function of the filling factor.
$u_\rho /K_\rho$ vanishes both for zero filling and half filling,
and in the latter case, the variation of $u_\rho/K_\rho$
is most dramatic for low $U$. If we now consider the phonon
coupling constant to be fixed at an arbitrary non--zero value, we
can get arbitrarily close to $b/u_\rho c=1$
by, for example, increasing the filling factor towards $n=1$.
As pointed out in Sec. \ref{Greens subsection},
when $u_\rho/K_\rho$ approaches from above the critical value
$u_\rho^*/K_\rho^*$ defined in Eq. (\ref{critical}),
the quantity $A$ which appears in the Green function exponent
and in the SDW and CDW correlation function exponents {\it diverges},
and $B$ decreases to a finite value
$B^*\equiv u_\rho^*/(u_\rho^{*2}+c^2)^{1/2}$. Consequently, SDW (CDW)
fluctuations are suppressed as we approach the Wentzel--Bardeen
singularity, and the system is pushed towards the triplet
superconducting phase,
via the metallic phase depicted in Figs. \ref{fig3} and \ref{fig4}.
Given the fact that $B^*\propto g^2$ and
$1/2\leq K_\rho\leq 1$
the superconductivity criterion, $B\leq K_\rho$,
can be met for sufficiently small
$g\neq 0$, by approaching the Wentzel--Bardeen singularity.
The system can be driven in the superconducting phase
by approaching half filling.

\subsection{Extended Hubbard model in the limit
$U\rightarrow\infty$}
\label{Large U limit}

We now add a nearest neighbor interaction term:
\begin{equation}
V\sum_i n_in_{i+1}~,\label{extended Hubbard}
\end{equation}
to the electronic Hamiltonian of Eq. (\ref{Hubbard model}) and
let $U\rightarrow\infty$. Each site can be occupied at most by
one electron, so that the system reduces effectively to spinless
fermions with the substitution $k_F\rightarrow 2k_F$.
At half filling (which corresponds to $n=1/2$ here as the Fermi
momentum
is $2k_F$), the spinless fermion Hamiltonian
with nearest neighbor interaction can be mapped
to the Hamiltonian of the XXZ model for a spin chain. The latter
Hamiltonian can be solved exactly
with the Bethe ansatz \cite{Yang}. For this reason, it is possible
\cite{Schulz2,Shankar} to get analytic expressions for the parameters
$K_\rho$ and $u_\rho$:
\begin{mathletters}
\begin{eqnarray}
K_\rho&=&(2+4\sin^{-1}(v)/\pi)^{-1}\label{K quarter}\\
u_\rho&=&\pi t\sqrt{1-v^2}/cos^{-1}v~,\label{u quarter}
\end{eqnarray}\end{mathletters}
for $v\equiv V/2|t|<1$. The case $v>1$ represents a dimerized
insulating phase \cite{Schulz}. One can therefore study the behavior
of
the correlation function exponents as the insulating phase is
approached.

The phase diagram is plotted in Fig. \ref{fig6}.
Obviously, the $U\rightarrow \infty$ limit pushes the system
towards
the SDW phase, but at relatively low $V/2t$, a strong coupling to
phonons
$b/cu_\rho\approx 0.94$ destroys the
SDW fluctuations. As $V/2t$ increases from $0$ to $1$,
the effects of the phonons become less pronounced, and the metallic
region
gradually disappears.

\section{Application to electron wave guides}
\label{wave guide section}

The results described in the previous section can be applied
in a different context. The contribution of the
Hamiltonian in Sec. \ref{model section} which describes the
coupling between the charge and phonon fields can alternatively
represent the coupling between two charge fields associated
with two interacting quasi one--dimensional electron
strings corresponding to two
transverse modes in an electron wave guide.

The question of the stability of coupled one--dimensional interacting
electrons systems has
been the focus of recent interest \cite{Schulz2,Anderson,Wen}.
Ref. \cite{Anderson} argues that the coupled system is stable,
in the sense that it stays in the Luttinger liquid phase, as long
as the hopping matrix element between the two chains remains
below a
critical value. Alternatively, a renormalization group
argument proposed in Ref. \cite{Schulz2} seems to imply that the
coupled system is unstable for an arbitrarily small coupling. Both Ref.
\cite{Schulz2} and
Ref. \cite{Anderson} assume a transverse hopping matrix element
as the mechanism for destabilisation.

We reexamine this question from a slightly different starting point,
considering the electrostatic effects between the two chains.
We will show below that a threshold coupling
will induce superconducting fluctuations in one of the chains, and in
both
chains as the coupling is further increased.
After all, if the two electron chains can be associated with two
orthogonal transverse states, it is reasonable to assume
that in the absence
of impurities, the (screened)
Coulomb interaction between the two chains is the basic
interaction mechanism which breaks the orthogonality between the
two
transverse states. In this situation, Eq. (\ref{coupling}) represents
a local density--density interaction:
\begin{eqnarray}
V_{1-2}&=&\pi g\sum_{s,s^\prime=\uparrow,\downarrow}
\int dx~ \psi_{1s}^\dagger(x)\psi_{1s}(x)
\psi_{2s^\prime}^\dagger(x)\psi_{2s^\prime}(x)
\nonumber\\
&=&~g{\pi\over 2}\int dx (\partial_x\varphi_{\rho 1})
(\partial_x\varphi_{\rho 2})~,
\label{coupling L}
\end{eqnarray}
where $\psi_{is}$  is the electron field operator
associated with an electron on chain $i$ with spin $s$,
and $\varphi_{\rho i}$ is the charge bosonic field of the same chain.
Note that the last equality  of Eq. (\ref{coupling L})
follows from the fact that the two transverse modes have distinct
Fermi velocities. Hence the fast varying components of the density
operators in the first line of Eq. (\ref{coupling L}) do {\it not}
compensate each other as in the coupling between electrons
and $2k_F$ phonons.
In the case of quasi--one--dimensional systems created in a
GaAs/AlGaAs
heterostructure, the assumption of a short range interaction can be
justified because the surrounding metallic gates screen
all interactions. The interaction potential of Eq. (\ref{coupling L})
is precisely the same form of coupling which was recently used
\cite{Matveev} to study the transport properties of arrays of
one--dimensional chains.
This coupling  describes two effects:
first, the fact that the local ``site'' energy of, say, the first wire (1)
is raised (or lowered) in the presence of a nearby
density perturbation in the
second wire (2).
Secondly, Eq. (\ref{coupling L}) allows electrons from 1 and 2 to
{\it exchange} as a pair.

The remaining part of the Luttinger liquid Hamiltonian which
describes the uncoupled system is then:
\begin{equation}
H_{el}=H_{\rho 1}+H_{\sigma 1}+H_{\rho 2}+H_{\sigma 2}~,
\label{free couped chain}\end{equation}
with, for $\nu=\rho,\sigma$ and $i=1,2$
\begin{equation}
H_{\nu,i}={1\over 2}\int d x~\Biggl[u_{\nu i} K_{\nu i}\Pi_{\nu i}^2
+{u_{\nu i} \over K_{\nu i}}(\partial_x\varphi_{\nu i})^2\Biggr]~,
\label{nu i Hamiltonian}\end{equation}
with $\Pi_{\nu i}$ the canonical conjugate of $\varphi_{\nu i}$.
In general, the charge velocities associated with the two
electron chains can be different.

Using the results of Sec. \ref{Lagrangian section},
the effective propagator
describing the evolution in a given chain $i$ becomes:
\begin{equation}
D_{\rho i}({\bf k})=(K_{\rho i} u_{\rho i})^{-1}
\biggl(\omega^2+u_{\rho i}^2 k^2-{b^2k^4\over\omega^2+u_{\rho
j}^2k^2}
\biggr)~,
\label{propagator i}
\end{equation}
with $i\neq j$ and $b^2=g^2K_{\rho 1}K_{\rho 2}u_{\rho 1}u_{\rho
2}\hbar^2$.

The Green function exponent associated with electrons from chain $i$
then reads:
\begin{equation}
1+\delta_i={K_{\sigma i}\over 4}+{1\over 4K_{\sigma i}}+
{B_i\over 4K_{\rho i}}+{K_{\rho i}A_{ij}\over 4}~.
\label{Greens exponent i}
\end{equation}
where
\begin{mathletters}
\begin{eqnarray}
A_{ij}&=&{u_{\rho i}\over v_++v_-}\biggl(1+{u_{\rho j}^2\over
v_+v_-}\biggr)\label{A i j}\\
B_i&=&{u_{\rho i}\over v_++v_-}
\biggl(1+{v_+v_-\over u_{\rho i}^2}\biggr)\label{B i}\\
v_\pm&=&{1\over 2}(u_{\rho 1}^2+u_{\rho 2}^2)\pm{1\over 2}
\sqrt{(u_{\rho 1}^2-u_{\rho 2}^2)^2+4b^2}~.
\end{eqnarray}\end{mathletters}

The coupling between the two chains will favor superconducting
fluctuations in each chain, in analogy with the phonon case.
The correlations functions which provide the signature
for these fluctuations are read directly from Eqs. (\ref{singlet})
and (\ref{triplet}), with all fermions operators belonging to the same
chain.

The criterion
indicating strong superconducting fluctuations in a given chain $i$
is then translated
from Eqs. (\ref{criterion singlet}) and (\ref{criterion triplet}):
\begin{mathletters}
\begin{eqnarray}{B_i\over K_{\rho i}}+K_{\sigma i}&\leq&2~.~~~~~~~
{\rm (singlet)}
\label{criterion singlet i}\\
{B_i\over K_{\rho i}}+{1\over K_{\sigma i}}&\leq&2~.~~~~~~~{\rm
(triplet)}
\label{criterion triplet i}
\end{eqnarray}\end{mathletters}
It is important to note that this effect does not depend on
whether the interaction between the electron chains is positive
or negative. As a result, the strong superconducting fluctuations
induced by the interchain coupling could be observed
in wires where only two channels propagate, or alternatively
in one--dimensional coupled electron--hole systems.
We can also determine from our analysis of the phonon
case that the chain with the lower charge velocity
will be the first to go superconducting.

For completeness, it should be mentioned that there are two
additional
correlation functions describing superconducting fluctuations
for this particular system.
These occur when a singlet or triplet
Cooper pair is formed with one electron
from chain 1 and one electron from chain 2. We shall ignore
this additional superconductivity mechanism here, because the
inequalities (\ref{criterion singlet i}) and (\ref{criterion triplet i})
are in general satisfied for a lower threshold phonon coupling.

For specificity, we consider the case of two Hubbard chains coupled
by the interaction of Eq. (\ref{coupling L}):
we refer the reader to Sec. \ref{Hubbard section},
where the dimensionless coupling parameter has to be replaced
by $b/u_{\rho 1}u_{\rho 2}$.
Away from zero filling and half filling, we see that for a reasonably
small $U/t$, superconducting fluctuations will be induced
above a threshold coupling. Our result therefore
seems to be in qualitative agreement with Ref. \cite{Anderson}, where
the two chains, if initially in the Luttinger liquid regime, will
remain in this regime below the threshold coupling. Also, note that
in analogy to the electron--phonon case, the Wentzel--Bardeen
singularity can be exploited to drive the system in the
superconducting phase.

\section{Summary and Conclusion}
\label{conclusion section}

This work has focused on the low temperature properties
of a system of interacting electrons coupled
to phonons in one dimension. The low energy properties
of interacting electrons  can then be described with the
Luttinger liquid picture introduced by Haldane
\cite{Haldane} for spinless fermions, and
later on generalized by Schulz \cite{Schulz} for
electrons with spin.
When such a system is coupled to {\it low}
energy phonons with a deformation
potential coupling, the phonons can be integrated out,
allowing for an {\it exact} calculation of the correlation
function exponents as a function of the Luttinger liquid
parameters, the phonon coupling constant, and the sound velocity.

We first considered single particle properties, calculating
the Green function ``dressed'' by the phonon modes.
For a gas of free electrons, the coupling to
phonons {\it alone} is sufficient to introduce a singularity
of the momentum
distribution function at the Fermi surface,
as is the case for fermions in the presence
of an instantaneous, attractive interaction: the power law decay
of the Green function at large distances
has an exponent $1+\delta$
which
is always larger than $1$. For increasing, but small phonon coupling,
the
repulsive interaction further accentuates this effect. However,
beyond a threshold phonon coupling, $\delta$ has a minimum as
a function of $U$. This indicates that for strong enough phonon
coupling, at low $U$, the instantaneous repulsion and the retarded
attractive interaction work in opposite ways.

Next, we considered the many body aspects of the system. In two
and three
dimensions it is understood that the parameters
of the electron Hamiltonian can drive the system in a definite
``phase'' at vanishing temperature,
such as a SDW or CDW phase. In the presence of phonons,
a natural question is to ask whether the retarded attraction
mediated by the phonons can flip the system into a triplet
superconducting
phase.
In 2D and 3D, the interplay between
retarded attractive interaction and instantaneous repulsive interaction
cannot be accounted for in a precise matter. After all,
the BCS theory of superconductivity \cite{Schrieffer}
assumes from the start
an attractive interaction {\it without} retardation effects.
The interplay between
attractive retarded interaction and instantaneous repulsive interaction
can be resolved exactly in one dimension.
A set of inequalities was derived, which
determine which type of fluctuations are present in the system.

To illustrate our method, we applied our results to the
Hubbard model, where the Luttinger liquid parameters
can be determined through a numerical computation, for arbitrary
filling factor and on--site repulsion.
The SDW fluctuations  dominate at large $U$ and near zero
and half
filling, where correlation effects between electrons are known to
be important. The region in parameter space where triplet
superconducting
fluctuations dominate is separated
from the region of SDW fluctuations by a metallic region.
This constitutes a novelty, as previous attempts
\cite{Emery,Solyom}
to draw
a phase diagram for purely instantaneous interactions predicted
sharp transitions between the SDW (CDW) and superconducting regions.
The retarded interaction mediated by the phonons is
responsible for this new phase. For a fixed $U$, we
found that as the phonon coupling is increased,
superconducting fluctuations
are most likely to occur near quarter filling, or
away from half filling and zero filling where correlation effects
are most important.
A natural extension of our investigation would be to study other models
of interacting electrons with nearest neighbor interaction,
such as the extended Hubbard model, where, however,
it is necessary to rely on numerical estimates for the
Luttinger liquid parameters.

We noted that by varying the electron
filling factor, the Wentzel--Bardeen
singularity can be approached for arbitrary values of the
electron--phonon coupling. This singularity \cite{Wentzel,Bardeen}
was previously understood to be the point where the coupling
is so strong that the system acquires a negative compressibility.
Nevertheless, for {\it correlated} electrons systems with
on site repulsion such as the Hubbard model, the singularity
is reached by decreasing the charge velocity associated with the
particle--hole excitations. Near this point, SDW (CDW)
correlations are totally suppressed, and the system is pushed
through the metallic phase towards the superconducting
phase.
We finally note here that this behavior is reminiscent of a characteristic
feature of high temperature superconductivity materials: the increase
of doping (and thus change of filling factor) changes
the phase of the material from antiferromagnetic to
metallic and finally to superconducting (at low enough temperatures).
Since there is some indication that the 2D layers of these new materials
can be described in terms of Luttinger liquids\cite{Anderson},
it is not unreasonable
to conjecture that the mechanism discussed here could be of relevance
for these materials. However, this is an
open question which deserves further investigation.

Next, we argued that our results can be applied to study the
(in)stability of two electrons (or one electron--hole) modes
in an electron wave guide,
coupled with a short range
density--density interaction.
For specific Hamiltonians decribing the interaction between electrons
within the chains, we can determine
the threshold coupling where the chains become superconducting.
Alternatively, the transition to the superconducting regime
could be observed by varying the electron density, expoiting once
again the WB singularity.
This could be probed by studying the periodicity
of the persistent current of a two channel mesoscopic ring
as a function of electron density. Such rings
with few transverse channels
can now be
fabricated using GaAs/AlGaAs heterostructures with metallic gates,
as in the remarkable experiment of Mailly {\it et al.} \cite{Mailly}.
The electron density can be varied my means of an electrostatic gate.
The flux causes a twist
in the boundary conditions, but does not affect the bulk properties,
such as the correlation function exponents. In addition,
the mutual inductance of the two transverse modes can be
neglected because its effects are proportional to
the small ratio $\alpha v_F/c_l$, with $\alpha\sim 137$
the fine structure
constant and $c_l$ the
velocity of light \cite{Martin}.
In the metallic regime,
the power spectrum
has a peak associated with the flux quantum
$\phi_0=hc/e$, whereas in the superconducting
case, this peak should give place to a peak
at the {\it superconducting} flux quantum
$\phi_0/2$.
Finally,
the study of the stability of two coupled transverse channels
could in principle be generalized to an arbitrary
number of channels, as it was recently shown that the
correlation function exponents for the multicomponent
Tomonaga--Luttinger model can be obtained formally
\cite{Penc} using conformal invariance arguments.

There are still challenging problems to be addressed.
First, we have neglected throughout this paper the coupling between
the fast oscillating part of the electron operator to the
$2k_F$ phonon mode. This coupling is known to be at the origin
of the Peierls instability, but is absent for the case of the
coupled electron modes in the wave guide. For
electron--phonon systems, it would be interesting to study
whether  the dramatic effects triggered by the Wentzel--Bardeen
singularity can survive this additional interaction. This question
can be answered in the case where the electron bandwidth is comparable to
the phonon cutoff, where the retardation effects
associated with the $2k_F$ phonon processes can be
neglected, and this additional coupling amounts simply
to a reduction of the on site repulsion parameter. The
Wentzel--Bardeen singularity survives the $2k_F$ phonon
processes in this case. In most
physical systems, however,
the phonon cutoff is smaller than the electron cutoff,
and these retardation effects have to be taken into account
\cite{Voit1,Zimanyi}.

Second, the present treatment uses exact results for the isolated
Hubbard model,
and applies it to a system which is coupled to phonons.
It would be desirable to have a self--consistent treatment,
where the effects of the correlated electrons on the lattice
are also described \cite{Nozieres}.

\acknowledgements
We thank C.P. Enz, A. J. Leggett, and S. Trugman for useful
discussions. The work of D.L. is supported by NSERC of Canada.
\appendix{Time evolution}
\label{appendix time}

In order to calculate the commutator in Eq. (\ref{normal order
Greens}),
we need to find the explicit time dependence of the operators
$\varphi_\rho$, $\varphi_\sigma$, $\Pi_\rho$ and $\Pi_\sigma$.
The evolution in real time is determined by the following equation of
motions
($\hbar=1$):
\begin{mathletters}
\begin{eqnarray}
\partial_t\varphi_\nu&=&K_\nu u_\nu\Pi_\nu~,~~~\nu=\rho,\sigma
\label{evolution phi}\\
\partial_t\Pi_\sigma&=&{u_\sigma\over K_\sigma}\partial_x^2
\varphi_\sigma\label{evolution sigma}\\
\partial_t\Pi_\rho&=&{u_\rho\over
K_\rho}\partial_x^2\varphi_\rho
+g\partial_x^2d\label{evolution rho}\\
\partial_td&=&\zeta^{-1}\Pi_d\label{evolution d}\\
\partial_t\Pi_d&=&\zeta
c^2\partial_x^2d+g\partial_x^2\varphi_\rho~.
\label{evolution pi d}
\end{eqnarray}
\end{mathletters}
Assuming for the above quantities a dependence
in time and space of the form $\exp(ikx-i\omega t)$
we obtain the dispersion relations:
\begin{mathletters}
\begin{eqnarray}
\omega_{\rho,\pm}^2&=&v_\pm^2 k^2\label{dispersion rho}
\\ \omega_\sigma^2&=&u_\sigma^2k^2~,\label{dispersion sigma}
\end{eqnarray}
\end{mathletters}
for the charge and spin fields. The hybrid velocities
$v_\pm$ are defined in Eq.~(\ref{velocities}).

At $t=0$, the fields can be expressed as a Fourier series:
\begin{mathletters}
\begin{eqnarray}
\varphi_\rho(x,0)&=&\sum_k\biggl({1\over 2L|k|}\biggr)^{1/2}
e^{ikx}(a_k^\dagger+a_{-k})\label{t 0 rho}\\
\Pi_\rho(x,0)&=&i\sum_k\biggl({|k|\over 2L}\biggr)^{1/2}
e^{ikx}(a_k^\dagger-a_{-k})\label{t 0 pi rho}\\
\varphi_\sigma(x,0)&=&\sum_k\biggl({1\over 2L|k|}\biggr)^{1/2}
e^{ikx}(b_k^\dagger+b_{-k})\label{t 0 sigma}\\
\Pi_\sigma(x,0)&=&i\sum_k\biggl({|k|\over 2L}\biggr)^{1/2}
e^{ikx}(b_k^\dagger-b_{-k})\label{t 0 pi sigma}\\
d(x,0)&=&\sum_k\biggl({\hbar\over 2Lc|k|\zeta}\biggr)^{1/2}
e^{ikx}(c_k^\dagger+c_{-k})\label{t 0 d}\\
\Pi_d(x,0)&=&i\sum_k\biggl({\hbar c|k|\zeta\over 2L}\biggr)^{1/2}
e^{ikx}(c_k^\dagger-c_{-k})~,\label{t 0 pi d}
\end{eqnarray}
\end{mathletters}
where $a_k$ ($b_k$) is an annihilation operator characterizing a
charge
(spin) excitation at momentum $k$, and $c_k$ is a phonon
annihilation
operator. These expressions are implicitly regularized by the large
momentum cut-off, $exp{(-\epsilon |k|L/2\pi})$, and furthermore
the $k=0$ mode is excluded in the sum.

Eq. (\ref{dispersion rho}) indicates that the time dependence
of the charge fields should be of the form:
\begin{mathletters}
\begin{eqnarray}
\varphi_\rho(x,t)&=&\sum_k\biggl({1\over 2L|k|}\biggr)^{1/2}e^{ikx}
\nonumber\\
&~&\times \sum_{\alpha,\beta=\pm}A_{\alpha\beta}
e^{-\beta k v_\alpha t}
\label{time rho}\\
\Pi_\rho(x,t)&=&\sum_k\biggl({|k|\over 2L}\biggr)^{1/2}e^{ikx}
\nonumber\\
&~&\times \sum_{\alpha,\beta=\pm}B_{\alpha\beta}
e^{-\beta k v_\alpha t}
{}~.\label{time pi rho}
\end{eqnarray}
\end{mathletters}
Similarly, the time dependence of the spin fields is determined
from Eq. (\ref{dispersion sigma}):
\begin{mathletters}
\begin{eqnarray}
\varphi_\sigma(x,t)&=&\sum_k\biggl({1\over
2L|k|}\biggr)^{1/2}e^{ikx}
\biggl[C_+e^{-iku_\sigma t}+C_-e^{iku_\sigma t}\biggr]
\nonumber\\ \label{time sigma}\\
\Pi_\sigma(x,t)&=&\sum_k\biggl({|k|\over 2L}\biggr)^{1/2}e^{ikx}
\biggl[D_+e^{-iku_\sigma t}+D_-e^{iku_\sigma t}
\biggr]~.\nonumber\\ \label{time pi sigma}
\end{eqnarray}
\end{mathletters}
The operators $A_{\pm\pm}$, $B_{\pm\pm}$, $C_\pm$ and
$D_\pm$
are determined by expressing the initial values
$\varphi_\nu(x,0)$, $\partial_t\varphi_\nu(x,0)$,
$\partial_t^2\varphi_\nu(x,0)$, $\partial_t^3\varphi_\nu(x,0)$,
in terms of spatial derivatives via Eqs.~(\ref{evolution phi})-
(\ref{evolution pi d}),
and similarly for $\Pi_\nu$.
We then obtain:
\begin{mathletters}
\begin{eqnarray}
A_{\alpha\beta}(k)&=&{1\over 2(v_\alpha^2-v_{-\alpha}^2)}
\Biggl((u_\rho^2-v_{-\alpha}^2)\biggl[a_k^\dagger
\biggl(1-{\beta u_\rho K_\rho\over v_\alpha}{\rm
sgn}(k)\biggr)\nonumber\\
&~&~~~~~~~~~~~~~~~~+
a_{-k}\biggl(1+{\beta u_\rho K_\rho \over v_\alpha}{\rm
sgn}(k)\biggr)\biggr]
\nonumber\\&~&~~~~~~~~~~~~~~~~+iK_\rho u_\rho g
(c\zeta)^{-1/2}\biggl[c_k(-1+
{\beta c\over v_\alpha}\biggr)\nonumber\\
&~&~~~~~~~~~~~~~~~~+c_{-k}^\dagger\biggl(1+{\beta c\over
v_\alpha}{\rm sgn}(k)\biggr)\biggr]\Biggr)\label{A plus}\\
B_{\alpha\beta}(k)&=&{1\over 2(v_\alpha^2-v_{-\alpha}^2)}
\Biggl(ia_k^\dagger
\biggl(u_\rho^2-v_{-\alpha}^2-{\beta\over K_\rho v_\alpha u_\rho}
(u_\rho^4-u_\rho^2v_{-\alpha}^2+b^2){\rm
sgn}(k)\biggr)\nonumber\\
&~&~~~~~~~~~~~~~~~~
+ia_{-k}\biggl(-u_\rho^2+v_{-\alpha}^2-{\beta\over K_\rho v_\alpha
u_\rho}
(u_\rho^4-u_\rho^2v_{-\alpha}^2+b^2){\rm sgn}(k)\biggr)
\nonumber\\&~&~~~~~~~~~~~~~~~~+g (c\zeta)^{-1/2}\biggl[c_k\biggl(-
c-
{\beta \over v_\alpha}(u_\rho^2+c^2-v_{-\alpha}^2){\rm
sgn}(k)\biggr)
\nonumber\\
&~&~~~~~~~~~~~~~~~~+c_{-k}^\dagger\biggl(-c+{\beta \over v_\alpha}
(u_\rho^2+c^2-v_{-\alpha}^2){\rm sgn}(k)\biggr)\biggr]\Biggr)
\label{B plus}\\
C_\alpha(k)&=&\biggl({1-\alpha K_\sigma {\rm sgn}(k)\over
2}\biggr)
b_k^\dagger
+\biggl({1+\alpha K_\sigma {\rm sgn}(k)\over 2}\biggr)b_{-
k}\label{C
plus}\\
D_\alpha(k)&=&i\biggl({1\over 2}-\alpha {{\rm sgn}(k)\over
2K_\sigma}\biggr)
b_k^\dagger
+\biggl(-{1\over 2}-\alpha {{\rm sgn}(k)\over
2K_\sigma}\biggr)b_{-k}
\label{D plus}
\end{eqnarray}\end{mathletters}
with $\alpha,\beta=\pm$.
These operators satisfy the commutation relations:
\begin{mathletters}
\begin{eqnarray}
\left[B_{\alpha\beta}(k),A_{\gamma\xi}(k^\prime)\right]
&=& -i\delta_{k,k^\prime}
\delta_{\alpha,\beta}\delta_{\gamma,\xi}\biggl({u^2-v_{-
\alpha}^2\over
v_\alpha^2-v_{-\alpha}^2}\biggr)\label{commutation A B}\\
\left[ D_\alpha (k),C_\beta(k^\prime)\right]&=& -i\delta_{k,
k^\prime}
\delta_{\alpha,\beta}\label{commutation D C}
\end{eqnarray}
\end{mathletters}
These relations are used in turn to calculate the commutators
which appear in the normal ordered products of Sec.
\ref{Greens subsection}, with the substitution $t\rightarrow i\tau$
for the thermodynamic quantities:
\begin{mathletters}
\begin{eqnarray}
\left[\theta_\rho(x,\tau),\varphi_\rho(0,0)\right]&=&
-{1\over 4\pi}\biggl({u_\rho^2-v_-^2\over v_+^2+v_-^2}
\ln\biggl[{1-\bar{z}(x-v_+i\tau)\over 1-z(x-v_+i\tau)}
{1-\bar{z}(x+v_+i\tau)\over 1-z(x+v_+i\tau)}\biggr]\nonumber\\
&~&~~~~~~+{u_\rho^2-v_+^2\over v_-^2+v_+^2}
\ln\biggl[{1-\bar{z}(x-v_-i\tau)\over 1-z(x-v_-i\tau)}
{1-\bar{z}(x+v_-i\tau)\over 1-z(x+v_-i\tau)}\biggr]\biggr)
\nonumber\\
\label{commutation rho}\\
\left[\theta_\sigma(x,\tau),\varphi_\sigma(0,0)\right]&=&
-{1\over 4\pi}\ln\biggl[{1-\bar{z}(x-u_\sigma i\tau)\over 1-z(x-
u_\sigma
i\tau)} {1-\bar{z}(x+u_\sigma i\tau)\over 1-z(x+u_\sigma
i\tau)}\biggr].\nonumber\\
\label{commutation sigma}
\end{eqnarray}
\end{mathletters}
Noting that
$\left[\theta_\nu(x,\tau),\varphi_\nu(0,0)\right]=
\left[\theta_\nu(0,0),\varphi_\nu (-x,-\tau)\right]$,
we obtain the explicit form for $g_n$.
If in addition $x\neq0$, we may put $\epsilon=0$ and use
that $\bar{z}(x+vi\tau)=1/{z}(x-v i\tau)$,
which then leads to the second equality in
Eq.(\ref{compact normal}) after some simple manipulations.

\appendix{Fourier sums}
In this Appendix we compute the Fourier sums which appear in Eq.
(\ref{Greens result}) in the limit $T\rightarrow 0$.
For that purpose it is convenient to rewrite
the quantities to be summed over in terms of irreducible fractions
expressed as functions of $\omega^2$ and $k^2$:
\begin{mathletters}
\begin{eqnarray}
D_\rho^{-1} ({\bf k})=u_\rho K_\rho &\Biggl(&
{c^2-v_+^2\over v_+^2-v_-^2}
\biggl({1\over \omega^2+v_-^2k^2}-{1\over \omega^2+v_+^2k^2}
\biggr)\nonumber\\
&~&+{1\over \omega^2+v_-^2k^2}\Biggr)\label{irreducible 1}\\
{u_\rho K_\rho D_\rho ({\bf k}) -\omega^2\over K_\rho
u_\rho^3k^2D_\rho({\bf k})}&=&
{c^2-b^2/u_\rho^2-v_+^2\over v_+^2-v_-^2}\biggl(
{1\over\omega^2+v_-^2k^2}-{1\over\omega^2+v_+^2k^2}\biggr)
\nonumber\\
&~&+{1\over \omega^2+v_-^2k^2}~,
\label{irreducible 2}\end{eqnarray}\end{mathletters}
where the velocities $v_\pm$ are defined in Eq. (\ref{velocities}).

With the above decompositions, the sums over Matsubara
frequencies
in Eq. (\ref{Greens result}) yield, in the $T\rightarrow 0$ limit:
\begin{mathletters}
\begin{eqnarray}
{1\over\beta}\sum_\omega D_\rho^{-1}({\bf k})&=&
{u_\rho K_\rho\over 2|k|(v_++v_-)}\biggl(1+{c^2\over v_+v_-}\biggr)
\label{Matsubara 1}\\
{1\over\beta}\sum_\omega D_\sigma^{-1}({\bf k})&=&{K_\sigma
\over 2 |k|}\label{Matsubara 2}\\
{1\over\beta}\sum_\omega{\omega\sin(\omega \tau)\over
D_\rho({\bf k})}&=&
{\rm sgn}(\tau){u_\rho K_\rho\over 2}\biggl({v_-^2-c^2\over
v_-^2-v_+^2}e^{-v_-|k \tau|}+{v_+^2-c^2\over
v_+^2-v_-^2}e^{-v_+|k\tau|}\biggr)
\nonumber\\ \label{Matsubara 3}\\
{1\over\beta}\sum_\omega
{\omega\sin(\omega \tau)\over D_\sigma({\bf k})}&=&{\rm
sgn}(\tau)
{u_\sigma K_\sigma\over 2}e^{-u_\sigma |k\tau|}
\label{Matsubara 4}\\
{1\over\beta}\sum_\omega{\cos(\omega \tau)\over
D_\rho({\bf k})}&=&{K_\rho\over 2|k|}\biggl({u_\rho\over v_-}
{v_-^2-c^2\over v_-^2-v_+^2}e^{-v_-|k\tau|}+{u_\rho\over v_+}
{v_+^2-c^2\over v_+^2-v_-^2}e^{-v_+|k\tau|}\biggr)
\nonumber\\ \label{Matsubara 5}\\
{1\over\beta}\sum_\omega{\cos(\omega \tau)\over D_\sigma({\bf
k})}&=&
{K_\sigma\over 2|k|}e^{-u_\sigma |k\tau|}~,
\label{Matsubara 6}\end{eqnarray}\end{mathletters}
where ${\rm sgn}(\tau)=1,0,-1$, if $\tau >$,$=$,$<0$.
Finally, the summations over momentum which occur in the
calculation of the Green function exponent are:
\begin{mathletters}
\begin{eqnarray}
{\pi\over L}\sum_k{1\over |k|}&=&-\ln\epsilon \label{Fourier 1}\\
{\pi\over L}\sum_k{\cos(kx)e^{-v|k\tau|}\over |k|}&=&-\ln|1-
z(x+iv|\tau|)|
\label{Fourier 2}\\
{\pi\over L}\sum_k{\sin(kx)e^{-v|k\tau|}\over k}&=&{1\over 2i}
\ln{1-\bar{z}(x+iv|\tau|)\over 1-z(x+iv|\tau|)}~,
\label{Fourier 3}
\end{eqnarray}\end{mathletters}
where $z(x)=\exp(2\pi ix/L-\epsilon)$, and where again we have
implicitly used the large momentum cut-off, $exp{(-\epsilon
|k|L/2\pi})$,
and omitted the $k=0$ term.

\figure{Green function exponent for the Hubbard model:
a) at fixed filling factor, as a function of $u=U/t$ for
$b/cu_\rho=0$ (solid line), $b/cu_\rho=0.2$ (dotted line),
$b/cu_\rho=0.4$ (dashed line) and $b/cu_\rho=0.6$ (dashed--dotted
line).
b) for fixed $U$, as a
function of filling factor and for the same values of $b/cu_\rho$ as in
Fig. 1a.\label{fig1}}
\figure{Phase diagram for $U=0.01 t$, for low filling factor
$n<0.2$, as a function of the phonon coupling $b/cu_\rho$, using the
perturbative result of the inequalities (\ref{weak S D W}) and
(\ref{weak triplet}). $SDW$, $M$, and $SC$ refer to the
spin density waves, metallic, and triplet superconducting regions.
\label{fig2}}
\figure{Phase diagram of the Hubbard model coupled to phonons,
as a function of filling factor and phonon coupling constant
$b/cu_\rho$:
a) $U/t=0.7$. b) $U/t=0.3$. c) $U/t=0.1$ .
We use the same labels as in Fig. \ref{fig2}. \label{fig3}}
\figure{Phase diagram of the Hubbard model coupled to phonons
at quarter filling ($n=1/2$), as a function of $U/t$ and $b/cu_\rho$.
\label{fig4}}
\figure{Plot of $u_\rho/K_\rho$ as a function of the filling factor $n$.
$u_\rho$ is measured in units of $v_F(n=1)/2=at$. From top to bottom:
$U/t=16,8,4,2$.
Note the abrupt change for small values of $U$ as one approaches half
filling.\label{fig5}}
\figure{Phase diagram for the extended Hubbard model in the limit
$U\rightarrow\infty$, as a function of the nearest neighbor
interaction $V/2t$ and the phonon coupling constant $b/cu_\rho$.
\label{fig6}}

\begin{references}
\bibitem{footnote0} This singularity was left unnoticed
in Ref. \cite{Voit0} because the
correlation function exponents given there appear to be incorrect.
\bibitem{footnote1}In contrast to $g_n$, the other oscillatory factors
in Eq.(\ref{end result Greens})
become unity, since
they should be evaluated at $\tau=0$ with  ${\rm sgn}(0)=0$ for the
calculation
of $<\psi^\dagger(x,0)\psi>$ (see Eqs.(\ref{Matsubara 3},
\ref{Matsubara 4})).
\end{references}
\end{document}